\documentclass[11pt]{article}

\usepackage{epsf}
\usepackage{color}
\usepackage{amsmath,amssymb,color,graphics,amscd,amsfonts}

\begin{document}

\begin{flushright} 
 June, 2012  \\
 OIQP-12-04
\end{flushright} 

\vspace{0.1cm}

\begin{Large}
\vspace{1cm}
\begin{center}
{\bf Correlation functions 
and representation bases 
in free ${\cal N}=4$ Super Yang-Mills} \\ 
\end{center}
\end{Large}

\vspace{0.2cm}

\begin{center}
{\large Yusuke Kimura}\footnote{
Okayama Institute for Quantum Physics, Kyoyama 1-9-1, Kita-ku, Okayama, 700-0015, Japan; 
londonmileend@gmail.com}
\\ 


\end{center}


\begin{abstract}
\noindent 

We study exact
correlation functions of ${\cal N}=4$ SYM at zero coupling. 
It has been known that 
it is convenient to label local gauge invariant operators by irreducible 
representations of symmetric groups/Brauer algebras.  
We first review the construction of representation bases 
from the viewpoint of the enhanced symmetry structure of the free theory.  
We present a basis of multi-matrix models using elements 
of Brauer algebras,
generalising our previous construction for two matrices. 
We will compute multi-point 
functions of the basis with the exact $N$-dependence. 
In particular we study three-point functions 
of a class of BPS operators, and we find that 
they are given  
by a branching rule of the Brauer algebra. 
The three-point functions take a factorised form 
if representations on the operators satisfy a relation.

\end{abstract}

\vspace{0.5cm}



\section{Introduction}

During the last decade, 
the planar limit of ${\cal N}=4$ Super Yang-Mills (SYM) has had a deep impact 
by the discovery of   
integrability, and 
the new structure has raised an expectation that  
${\cal N}=4$ SYM in the planar limit might be solved exactly. 
In contrast, 
we have a lack of knowledge about the non-planar theory. 
(See the papers \cite{1012.3982,1012.3997} for review.)

Study of ${\cal N}=4$ SYM in the free field limit 
is a direction 
to focus on non-planar corrections. 
Even in the free theory, summing up all non-planar diagrams 
is a highly non-trivial issue,  
especially when our concern is operators with 
scaling dimensions of ${\cal O}(N)$ or ${\cal O}(N^2)$. 
Such operators 
are considered to be dual to 
giant gravitons \cite{0003075,0008015,0008016} or 
general curved geometries \cite{0409174}. 
One can be motivated to study 
${\cal N}=4$ SYM at zero coupling  
by recalling the fact that 
it is dual to a string theory defined on 
a highly curved space according to AdS/CFT. 
(In fact there is a conjecture that the free SYM is dual to 
a higher-spin gauge theory \cite{0002189,0103247,0205131,0405057}.)
Because the free field theory is much easier than 
the full theory, 
it would be a good lesson to study it  
in order to learn how to describe 
degrees of freedom corresponding to 
D-branes or geometries 
before going to the full interacting theory.

In this paper, we will compute exact correlation functions of local 
gauge invariant operators 
in ${\cal N}=4$ SYM at zero coupling utilising 
the recently developed group theoretic technique. 
With the help of group theory and representation theory, 
we have an efficient way 
to diagonalise 
two-point functions and obtain the exact $N$-dependence. 
The first result employing group theory 
was given in \cite{0111222, 0205221} in which exact 
correlators in the 1/2 BPS sector was computed. 
Later it was generalised to include more fields in 
\cite{0709.2158,0711.0176,0801.2061,0806.1911,0807.3696,0910.2170}. 
These studies showed that 
operators that diagonalise two-point functions 
are labelled by representations of groups or algebras.  
We call such bases representation bases. 

The organisation of this paper is as follows. 
In the next section 
we will explain how group representation theory can be 
an efficient tool in the free theory by reviewing the 
integrable 
structure of the free theory 
from the viewpoint of an enhanced symmetry 
given in \cite{0807.3696}.  
In section 
\ref{sec:Brauerbasis}, we will give an orthogonal 
basis of multi-matrix operators using the Brauer algebra, 
generalising our previous construction \cite{0709.2158} 
for two matrices.
We will also present explicit forms of the operator in some special sectors, 
which will be helpful to 
take a general meaning of 
representation labels 
on the operator.  
In section 
\ref{sec:correlation_functions}, we compute multi-point correlation functions 
of the Brauer operators. 
Section
\ref{correlator_BPS} is given for a study of correlation functions 
of the operators labelled by only
irreducible representations of the Brauer algebra. 
Such operators 
were shown to be BPS in our previous paper \cite{1002.2424}.
We will see that the three-point functions of the BPS operators 
are given by a branching rule of the Brauer algebra.  
The Brauer basis would be suitable for 
constructing a composite from two kinds of operators. 
We will find that certain multi-point functions take a factorised form 
when representation labels of the operators satisfy a relation.    
In section  
\ref{sec:discussions} we will discuss future directions. 
In some appendices, we give some detailed computations, 
a brief review of representation theory, and the construction of 
commuting higher charges 
of the free system.


\section{The free theory and higher conserved charges}
\label{sec:review_freetheory}

In this section, 
we review a symmetry structure of ${\cal N}=4$ 
SYM  at zero coupling, and then we give an introduction of 
constructing  
representation bases by regarding the symmetry structure as a guiding principle. 

Start from ${\cal N}=4$ 
SYM theory defined on $R^4$ with $U(N)$ gauge group.
For simplicity, we will discuss only the $SO(6)$
scalar sector, and 
the generalisation to the full sector 
will be mentioned in section  
\ref{sec:discussions}. 
It is convenient to combine the 
six scalar fields into 
the complex combinations
\begin{eqnarray}
X_{1}=\frac{1}{\sqrt{2}}(\Phi_1+i\Phi_2),\quad
X_{2}=\frac{1}{\sqrt{2}}(\Phi_3+i\Phi_4),\quad
X_{3}=\frac{1}{\sqrt{2}}(\Phi_5+i\Phi_6).
\end{eqnarray}
We also denote them by 
$X$, $Y$, $Z$. 

The two-point functions take the following form
\begin{eqnarray}
\langle O_{i}(x)O_{j}(y)\rangle _{SYM}
=\frac{c_{ij}(N)}{(x-y)^{d_i+d_j}}\delta_{d_i, d_j}, 
\end{eqnarray}
where $d_i$ is the classical dimension of 
a gauge invariant operator $O_i$. 
General gauge invariant operators 
are given by a product of an arbitrary number of 
single trace operators built from 
the fundamental fields $X_a$ and $X_a^{\dagger}$. 
At zero coupling, the scaling dimension is 
counted by the number of fields involved in the operator. 
The non-trivial 
$N$-dependence 
is encoded only in 
$c_{ij}$, 
which 
can be computed by 
the matrix integral with the Gaussian weight 
\cite{0205033}
\begin{eqnarray}
c_{ij}(N)=
\langle O_{i}O_{j}\rangle 
:=
\int \prod_a [dX_a dX_a^{\dagger}] 
e^{-2 tr(X_aX_a^{\dagger})}
O_{i}O_{j},
\end{eqnarray}
where 
the measure is normalised to give 
$\langle  X^i_{j}X^{\dagger}{}^k_l
\rangle =
\delta^i_{l}\delta_{j}^{k}$, and 
the fields are regarded 
as matrices without the space dependence. 
Throughout this paper, 
we do not focus on the space-time dependence 
because it is trivially recovered by the conformal invariance. 
In this way, 
the $N$-dependence of correlation functions 
in the free theory
can be captured by the 
correlation functions of the matrix model.
When an operator 
contains 
both $X_a$ and $X_a^{\dagger}$, 
we define the operator by removing self-contractions. 
We introduce the normal ordering symbol $:O:$ 
in order to denote that it is 
free from self-contractions.  

\quad 

We next 
consider the radial quantisation of the theory 
by the conformal transformation from 
$R^4$ to $R\times S^3$. 
The action of the 
scalar fields obtains a mass term due to 
the conformal coupling to the curvature of 
$S^3$. 
When our interest is the singlet sector under $SO(4)$ corresponding to
the isometry of $S^3$, 
we are left with the s-wave mode
of the spherical harmonics of $S^3$.  
The action is then given by the matrix quantum mechanics 
\begin{eqnarray}
S= \int dt \sum_a
  \mathrm{tr}
\left(
\frac{1}{2}\dot{X}_{a}\dot{X}_{a}^{\dagger}
-\frac{1}{2}X_{a}X_{a}^{\dagger}
\right).
\end{eqnarray} 
Defining annihilation and creation operators as we usually do,
we get 
a collection of harmonic oscillators 
\begin{eqnarray}
H=\sum_a tr(A_a^{\dagger}A_a+B_a^{\dagger}B_a),
\end{eqnarray} 
where 
we have ignored the zero-point energy. 
The Cartan elements of R-charge are given by 
\begin{eqnarray}
J_a=tr(-A_a^{\dagger}A_a+B_a^{\dagger}B_a).
\end{eqnarray} 
By the radial quantisation, the dilatation operator 
defined on $R^4$ is mapped to the Hamiltonian 
on $R\times S^3$. 
The only non-trivial commutation relations are 
\begin{eqnarray}
[A_{a}{}^i{}_{j},A_{b}^{\dagger}{}^k{}_{l}]=\delta_{ab}\delta_{j}^{k}\delta^{i}_{l}, 
\quad
[B_{a}{}^i{}_{j},B_{b}^{\dagger}{}^k{}_{l}]=\delta_{ab}\delta_{j}^{k}\delta^{i}_{l}.
\label{commutator_oscillator}
\end{eqnarray}
The $A$-operators commute with the $B$-operators.
States are built by acting with 
$A_{a}^{\dagger}$ and $B_{a}^{\dagger}$ 
on the vacuum uniquely determined by 
$A_{a}{}^i{}_{j} |0\rangle =0$ and $B_{a}{}^i{}_{j}|0\rangle =0$. 
Some descriptions of 
the matrix quantum mechanics 
in the 1/2 BPS sector 
were studied in \cite{0403110}. 

The matrix integrals can be 
obtained by introducing 
the coherent state 
\begin{eqnarray}
|X,X^{\dagger}\rangle =
:e^{trX_a^{\dagger} B_a ^{\dagger}}
e^{trX_a A_a ^{\dagger}}:
 | 0\rangle .  
\end{eqnarray}
For gauge invariant states $|\Psi_i\rangle $, 
the inner product can be 
expressed by the matrix integral 
as 
\begin{eqnarray}
\langle \Psi_i |\Psi_j\rangle =
\int \prod_a[dX_a dX_a^{\dagger}]e^{-2tr(X_aX_a^{\dagger})}
O_i(X,X^{\dagger})^{\dagger}O_j(X,X^{\dagger}), 
\label{ComplexMultimatrixModel}
\end{eqnarray} 
where 
$O_i$ is the gauge invariant operator corresponding to 
a state $|\Psi_i\rangle$ 
\begin{eqnarray}
O_i(X,X^{\dagger})=
\langle X,X^{\dagger}|\Psi_i\rangle .
\end{eqnarray} 
We note again that 
operators are defined to be free from  
self-contractions 
in the integral. 
The matrix quantum mechanics of the non-holomorphic sector 
is also reviewed in \cite{0911.4408} using a slightly 
different notation from here.

\quad


The system is a collection of the harmonic oscillators. 
Because it is an integrable system, it is expected to have 
a set of commuting conserved charges.  
We shall give a review of the construction 
of conserved charges 
based on \cite{0807.3696}. 
In the course of the construction, 
group representation theory is naturally introduced, and 
higher charges will be
simultaneously diagonalised by representation bases.

We now define 
\begin{eqnarray}
(G_{La}^B)^i{}_{j}:=(B_{a}^{\dagger}B_a)^i{}_{j}, 
\quad 
(G_{Ra}^B)^i{}_{j}
:=-B_{a}^{\dagger}{}^k{}_{j}B_{a}{}^i{}_{k}.
\label{building_blocks}
\end{eqnarray}
We find that 
\begin{eqnarray}
&&
(G_{La}^B)^k{}_{l}
B_{b}^{\dagger}{}^i{}_{j}|0\rangle 
=\delta_{ab}
B_{b}^{\dagger}{}^k{}_{j}\delta^i{}_{l}|0\rangle ,
\nonumber 
\\
&&
(G_{Ra}^B)^k{}_{l}B_{b}^{\dagger}{}^i{}_{j}
|0\rangle 
=-\delta_{ab}
B_{b}^{\dagger}{}^i{}_{l}\delta^k{}_{j}|0\rangle .
\label{u(N)actions_leftright}
\end{eqnarray}
In the first equation, the lower index of the $(B^{\dagger})^i{}_{j}$ is 
inert, while the upper index is inert in the second equation. 
The following commutators can be easily 
obtained from the basic commutator in (\ref{commutator_oscillator})
\begin{eqnarray}
&&[G_{La}^B,G_{Rb}^B]=0, 
\nonumber \\
&&[(G_{sa}^B)^{i}{}_{j},(G_{sb}^B)^{k}{}_{l}]=
\delta_{ab}
(
(G_{sa}^B)^{i}{}_{l}\delta_{j}^{k}
-(G_{sa}^B)^{k}{}_{j}\delta^{i}_{l}
)
\quad (s=L,R).
\label{uNcommutationrelation}
\end{eqnarray}
The above is six copies of $u(N)$ algebra. 
The usual adjoint $u(N)$
transformation is generated by $G_{La}+G_{Ra}$. 
The $A$-operators satisfy similar relations.

In terms of the generators, 
the Hamiltonian can be written by  
\begin{eqnarray}
H=\sum_a tr(G_{La}^{A}+G_{La}^{B}).
\end{eqnarray}
Because the Hamiltonian is a number operator, 
any operators built from $G_{La}$ and $G_{Ra}$ 
commute with the Hamiltonian. 
We call such operators conserved operators.
We will build 
commuting conserved charges 
by regarding the pieces in (\ref{building_blocks}) as 
fundamental building blocks. 

\quad

We shall first consider the system in which  
states are excited by $B_1^{\dagger}$ alone. It is the 1/2 BPS sector. 
We can find
\begin{eqnarray}
[H_{p},H_q]=0,\quad H_p:=
tr((B_1^{\dagger}B_1)^p)
=tr((G_{L1}^B)^p),
\label{conserved_on_schur}
\end{eqnarray}
where $p$, $q$ are positive integers. (See also \cite{0111222,0607033}). 
We have another sequence of conserved charges given by 
$tr((G_{R1}^B)^p)$, 
which have the same eigenvalues the operators 
$tr((G_{L1}^B)^p)$ have
(see (\ref{higher_charges_schur})).

The system we next consider is the system that contains $B_a^{\dagger}$ $(a=1,2,3)$.  
In this case, 
the algebra of conserved operators is rather
complicated, but we easily 
find that \footnote{
We also have $tr((G_{R1}^B)^p)$, 
$tr((G_{R2}^B)^p)$, $tr((G_{R3}^B)^p)$, 
which we do not show explicitly.  
}  
\begin{eqnarray}
tr((G_{L1}^B)^p), \quad tr((G_{L2}^B)^p),\quad tr((G_{L3}^B)^p)
\label{charges_C}
\end{eqnarray}
commute each other. 
We also find that any conserved charges 
built from the pieces in 
(\ref{building_blocks}) commute with the charges in (\ref{charges_C}). 
(The formula 
(\ref{formula_charges_commutator})
can be available to very this.)
Hence the charges in (\ref{charges_C}) 
are always included in a set of commuting conserved operators. 
We denote the set of the operators in 
(\ref{charges_C})
by $S$. 
In constructing commuting conserved charges,  
it is convenient to 
start from one of the following charges as well as $S$: 
\begin{eqnarray}
&& 
tr((G_{L1}^B+G_{L2}^B+G_{L3}^B)^p), \quad  
tr((G_{L1}^B+G_{L2}^B+G_{R3}^B)^p), \quad 
tr((G_{R1}^B+G_{L2}^B+G_{R3}^B)^p), \quad \nonumber \\
&&  
tr((G_{L1}^B+G_{R2}^B+G_{R3}^B)^p).  
\label{basic_charges}
\end{eqnarray}
Because they do not commute each other, 
we cannot put them together in a set.
Based on one of the charges above with $S$, we can 
form a set of commuting conserved charges. 

In what follows we will show how the basic conserved charges - 
$S$ and (\ref{basic_charges}) are diagonalised by representation theory. 
The basic ideas 
we shall exploit are as follows \cite{0807.3696}. 
From  
(\ref{u(N)actions_leftright}), 
these higher charges are nothing but Casimir operators 
associated with the $u(N)$ actions. 
We will use the group theoretic fact that 
Casimirs are constant on an irreducible 
subspace of $V^{\otimes m}\otimes \bar{V}^{\otimes n}$, where $V$ 
and $\bar{V}$ are the fundamental 
representation and anti-fundamental representation 
of $U(N)$.  

The construction of each set of commuting charges 
is studied more fully  
in appendix 
\ref{sec:algebra_conservedcharge}, 
and the eigenvalues are shown in 
appendix 
\ref{Casimirs_action}. 
Allowing excitations by $A^{\dagger}_a$ is a straightforward 
extension.

\quad 

We will construct a basis of  
gauge invariant states built from 
$n_1$ $B_1^{\dagger}$'s, $n_2$ $B_2^{\dagger}$'s, and  
$n_3$ $B_3^{\dagger}$'s. 
Start from the tensor product 
\begin{eqnarray}
B_1^{\dagger}{}^{\otimes n_1}\otimes 
B_2^{\dagger}{}^{\otimes n_2}\otimes
B_3^{\dagger}{}^{\otimes n_3}  
\end{eqnarray}
as an endomorphism on $V^{\otimes n_1+n_2+n_3}$. 
It is known that the tensor product representation 
is reducible
and  
can be decomposed as 
\begin{eqnarray}
V^{\otimes n_1+n_2+n_3}
&=&\bigoplus_{R \vdash (n_1+n_2+n_3)}
V_{R}^{U(N)}\otimes V_{R}^{S_{n_1+n_2+n_3}}.
\label{tensorproduct_space}
\end{eqnarray}
The expression $R \vdash n$ 
indicates that $R$ is summed over Young diagrams 
with $n$ boxes. 
$V_{R}^{U(N)}$ and $V_{R}^{S_{n_1+n_2+n_3}}$ 
are the vector space of an irreducible representation $R$
of $U(N)$ and that of $S_{n_1+n_2+n_3}$. 
The equation is known as Schur-Weyl duality. 
The Casimir operator 
$tr((G_{L1}+G_{L2}+G_{L3})^p)$ has a constant eigenvalue 
on the irreducible subspaces. 

We can decompose the irreducible representation of $S_{n_1+n_2+n_3}$ 
into irreducible representations of 
the subgroup 
$S_{n_1}\times S_{n_2}\times S_{n_3}$ as 
\begin{eqnarray}
V_{R}^{S_{n_1+n_2+n_3}}
=\bigoplus_{  
r_1\vdash n_1,r_2\vdash n_2,
r_3\vdash n_3} 
V_{R\rightarrow \vec{r}}
\otimes 
V_{\vec{r}}^{S_{n_1}\times S_{n_2}\times S_{n_3}},
\label{decomposeRinto_r}
\end{eqnarray}
where $\vec{r}:=(r_1,r_2,r_3)$. 
The dimension of $V_{R\rightarrow \vec{r}}$ is given by 
the number of times the representation 
$\vec{r}$ appears 
in the restriction of the representation 
$R$ of $S_{n_1+n_2+n_3}$ to $S_{n_1}\times S_{n_2}\times S_{n_3}$, 
which is expressed by the Littlewood-Richardson coefficient as
\begin{eqnarray}
Dim(V_{R\rightarrow \vec{r}})=g(\vec{r};R):=g(r_1,r_2,r_3;R). 
\end{eqnarray}
 
Let us take
a particular class of operators 
in the symmetric group 
$S_{n_1+n_2+n_3}$ 
that project on representations $R,\vec{r}$. 
We will denote it by $P^R_{\vec{r},ij}$. 
The indices $i,j$ are multiplicity indices running over 
$1,2,\cdots,g(\vec{r};R)$.
The important equations 
the operators satisfy
are \cite{0801.2061}
\begin{eqnarray}
&&
h P^{R}_{\vec{r},ij}h^{-1}=P^{R}_{\vec{r},ij} \quad 
(h \in S_{n_1}\times S_{n_2}\times S_{n_3}), 
\nonumber \\
&&
P^{R}_{\vec{r},ij}
P^{R^{\prime}}_{\vec{r}^{\prime},i^{\prime}j^{\prime}}
=
\delta^{R R^{\prime}}
\delta^{\vec{r}\vec{r}^{\prime}}
\delta_{ji^{\prime}}
P^{R}_{\vec{r},ij^{\prime}},
\nonumber \\
&& 
tr_{n_1+n_2+n_3}(P^{R}_{\vec{r},ij})=d_{\vec{r}} DimR \delta_{ij},
\end{eqnarray}
where $tr_{n_1+n_2+n_3}$ is a trace 
over the tensor product space 
(\ref{tensorproduct_space}). 
$d_{\vec{r}}$ is the dimension of 
an irreducible representation $\vec{r}$ of 
$S_{n_1}\times S_{n_2}\times S_{n_3}$,  
and $DimR$ is the dimension of 
an irreducible representation $R$ of $U(N)$. 
Making use of the operator 
$P^R_{\vec{r},ij}$, 
we can 
build a gauge invariant state 
\begin{eqnarray}
|R,\vec{r},ij \rangle :=
tr_{n_1+n_2+n_3}(
P^R_{\vec{r},ij}
B_1^{\dagger \otimes n_1}\otimes 
B_2^{\dagger \otimes n_2}\otimes
B_3^{\dagger \otimes n_3} 
) |0\rangle.
\end{eqnarray}
This basis is called restricted Schur basis. 
It forms an orthogonal basis 
\cite{0801.2061,0805.3025,0810.4217}. 
It was originally introduced to describe open string excitations on 
giant gravitons \cite{0411205,0701066,0710.5372}. 
For the action of the conserved charges on this basis, see 
Appendices \ref{sec:algebra_conservedcharge} and \ref{Casimirs_action}.

\quad 

We next focus on another set of conserved charges 
that includes 
$S$ and 
$tr((G_{L1}+G_{L2}+G_{R3})^p)$. 
This time 
it is convenient to start with the tensor product 
\begin{eqnarray}
B_1^{\dagger}{}^{\otimes n_1}\otimes 
B_2^{\dagger}{}^{\otimes n_2}\otimes
B_3^{\dagger}{}^{T \otimes n_3}  .
\label{tensorproduct_Brauer}
\end{eqnarray}
The tensor product representation 
can be decomposed as 
\begin{eqnarray}
V^{\otimes n_1+n_2}
\otimes \bar{V}^{\otimes n_3}
=\bigoplus_{\gamma }
V_{\gamma}^{U(N)}
\otimes V_{\gamma}^{B_N(n_1+n_2,n_3)},
\label{SWdual_Brauer}
\end{eqnarray}
where $\gamma$ runs over irreducible representations of 
the Brauer algebra $B_N(n_1+n_2,n_3)$. 
The irreducible representations are labelled by a pair of Young diagrams 
$(\gamma_+,\gamma_-)$ that have 
$n_1+n_2-k$ and $n_3-k$ boxes, where $k$ is an integer that satisfies 
$0\le k \le min(n_1+n_2,n_3)$. 
We also denote them 
by $(\gamma_+,\gamma_-,k)$ to 
place an importance 
on the value of 
$k$, 
because the integer $k$ plays an important role in this paper. 
The representation theory is briefly summarised in 
appendix \ref{sec:irreps_Brauer}. 
The Brauer algebra plays the same role the symmetric group 
does in Schur-Weyl duality (\ref{tensorproduct_space}). 

Because the Brauer algebra contains the group algebra 
$\mathbb{C}[
S_{n_1}\times S_{n_2}\times S_{n_3}]$ as a subalgebra, 
we have 
\begin{eqnarray}
V_{\gamma}^{B_N(n_1+n_2,n_3)}
=\bigoplus_{  
r_1\vdash n_1,r_2\vdash n_2,
r_3\vdash n_3
}
V_{\gamma \rightarrow r}\otimes 
V_{\vec{r}}^{
\mathbb{C}[
S_{n_1}\times S_{n_2}\times S_{n_3}]}, 
\label{Brauer_restriction_symmetric}
\end{eqnarray}
where $V_{\gamma \rightarrow \vec{r}}$ 
is a vector space associated with the restriction, whose 
dimension is given by 
\footnote{
It can be derived by decomposing the multiplicity as 
\begin{eqnarray}
M^{\gamma}_{\vec{r}}=\sum_{t\vdash (n_1+n_2)}
M^{\gamma}_{(t,r_3)}
M^{t}_{(r_1,r_2)},
\end{eqnarray}
where the two factors in the RHS come from the restrictions 
$B_N(n_1+n_2,n_3)\rightarrow 
\mathbb{C}[S_{n_1+n_2}]\times \mathbb{C}[S_{n_3}]$ 
and 
$\mathbb{C}[S_{n_1+n_2}]\rightarrow 
\mathbb{C}[S_{n_1}]\times \mathbb{C}[S_{n_2}]$. 
The multiplicity of the first restriction is given by the form 
(\ref{multiplicity_Brauer_symmetric}). 
We have another way to decompose the multiplicity 
\begin{eqnarray}
M^{\gamma}_{\vec{r}}=
\sum_{\gamma_1}
M^{\gamma}_{(\gamma_1,r_2)}
M^{\gamma_1}_{(r_1,r_3)},
\end{eqnarray}
where $\gamma_1$ is an irreducible representation of the $B_{N}(n_1,n_3)$. 
The two restrictions are 
$B_N(n_1+n_2,n_3)\rightarrow 
B_N(n_1,n_3)\times B_N(n_2,0)$ and 
$B_N(n_1,n_3)\rightarrow 
\mathbb{C}[S_{n_1}]\times \mathbb{C}[S_{n_3}]$. 
The multiplicity associated with the 
first one is referred to 
(\ref{Brauer_branching}) and (\ref{branching_n2=0}).
}
\begin{eqnarray}
M^{\gamma}_{\vec{r}}=
Dim(V_{\gamma \rightarrow \vec{r}})
=\sum_{t\vdash (n_1+n_2)}
\sum_{\tau\vdash k}g(\gamma_{-},\tau;r_3)
g(\gamma_{+},\tau;t)
g(r_1,r_2;t).
\label{multiplicity_brauerop}
\end{eqnarray}

We now introduce 
particular linear combinations of 
elements in the Brauer algebra
that are associated with representations 
$\gamma, \vec{r}$. We denote it by $Q^{\gamma}_{\vec{r},ij}$. 
(More information about it will be given in the next section.)
Making use of this operator, we construct a gauge invariant state as
\begin{eqnarray}
|\gamma,\vec{r},ij \rangle :=
tr_{n_1+n_2,n_3}(
Q^{\gamma}_{\vec{r},ij}
B_1^{\dagger \otimes n_1}\otimes 
B_2^{\dagger \otimes n_2}\otimes
B_3^{\dagger T \otimes n_3} 
)|0\rangle , 
\label{Brauer_state}
\end{eqnarray}
where the trace $tr_{n_1+n_2,n_3}$ is taken over 
the space (\ref{SWdual_Brauer}).
We call this basis Brauer basis or basis based on the Brauer algebra 
$B_N(n_1+n_2,n_3)$. When $n_2=0$, it is the basis
proposed in \cite{0709.2158,0807.3696}. 
In the next section 
we will supplement this basis with some properties. 
The conserved charges are examined more fully in  
Appendices \ref{sec:algebra_conservedcharge} and \ref{Casimirs_action}.

It is 
also possible to consider in stead of (\ref{tensorproduct_Brauer})
\begin{eqnarray}
B_1^{\dagger}{}^{\otimes n_1}\otimes 
B_2^{\dagger}{}^{T\otimes n_2}\otimes
B_3^{\dagger}{}^{ \otimes n_3} 
\end{eqnarray}
or 
\begin{eqnarray}
B_1^{\dagger}{}^{T\otimes n_1}\otimes 
B_2^{\dagger}{}^{\otimes n_2}\otimes
B_3^{\dagger}{}^{\otimes n_3},  
\end{eqnarray}
where $B_N(n_1+n_3,n_2)$ or 
$B_N(n_2+n_3,n_1)$ plays the role respectively. 
These are related to the conserved charges 
$tr((G_{L1}+G_{R2}+G_{L3})^p)$ 
or 
$tr((G_{R1}+G_{L2}+G_{L3})^p)$.  
These are completely similar to the case in which $B_N(n_1+n_2,n_3)$ plays the role.

\quad

We now give an expression of 
the conserved charges 
as differential operators on the matrices $X$, $Y$, etc.  
The operators in  
the set $S$ are mapped to
\begin{eqnarray}
tr((X\partial_X)^p), \quad tr((Y\partial_Y)^p),
\quad tr((Z\partial_Z)^p),
\end{eqnarray}
while the operators 
$tr((G_{L1}^B+G_{L2}^B+G_{L3}^B)^p)$ and   
$tr((G_{L1}^B+G_{L2}^B+G_{R3}^B)^p)$ are mapped to 
\begin{eqnarray}
tr((X\partial_X+Y\partial_Y
+Z\partial_Z)^p), \quad 
tr((X\partial_X+Y\partial_Y
-:\partial_Z Z:)^p), 
\end{eqnarray}
where 
$:(\partial_Z Z)_{ij}:=Z_{kj}(\partial_Z)_{ik}$. 
We also have 
$tr((G_{R1}^A)^p)\rightarrow tr((:\partial_{\bar{X}} \bar{X}:)^p)$ etc.

\quad 

Before moving to the next section, 
mentioned is   
the use of 
(\ref{building_blocks}) as a building-block of conserved quantities. 
It was possible to add a more general conserved building block like 
$(B_{a}^{\dagger})_{ij}(B_a)_{kl}$ in the set of building blocks. 
When we think about 
states excited by only $(B_{1}^{\dagger})_{ij}$, 
we only get the same set 
as (\ref{conserved_on_schur}). 
On the other hand, when we have more than one matrix, 
we can construct 
a different set of 
conserved operators 
if we are allowed to use $(B_{a}^{\dagger})_{ij}(B_a)_{kl}$
in addition to (\ref{building_blocks}). 
In fact, it was shown in \cite{0807.3696} that 
the building block 
$(B_{a}^{\dagger})_{ij}(B_a)_{kl}$ 
is needed to construct conserved charges 
to measure the labels of the basis given in 
\cite{0711.0176,0806.1911}.
We leave it as a future homework to 
give a complete classification of 
conserved charges by beginning with more general 
building blocks.


\section{Orthogonal basis for multi-matrix  
using Brauer algebra}
\label{sec:Brauerbasis}

In this section, we will study 
the representation basis 
(\ref{Brauer_state}) in more detail. 
This basis itself is new, but 
most properties are 
generalisations of 
the Brauer basis for two matrices given 
in \cite{0709.2158,0807.3696,0911.4408}.

Recall the basis
\begin{eqnarray}
O^{\gamma}_{\vec{r},ij}
(X,Y,Z)
:=
\langle X,Y,Z|\gamma,\vec{r},ij \rangle =
tr_{n_1+n_2,n_3}(
Q^{\gamma}_{\vec{r},ij}
X^{\otimes n_1}\otimes 
Y^{\otimes n_2}\otimes
Z^{T \otimes n_3} 
).
\end{eqnarray}
The operator $Q^{\gamma}_{\vec{r},ij}$ 
is a linear combination of elements in the Brauer algebra \cite{0709.2158,0807.3696}, 
\begin{eqnarray}
Q^{\gamma}_{\vec{r},ij}=Dim\gamma 
\sum_{b\in B_{N}(n_1+n_2,n_3)}
\chi^{\gamma}_{\vec{r},ji}(b)b^{\ast}.
\end{eqnarray}
Here we have defined 
the restricted character\footnote{
If we take the sum of $\vec{r}$ and $i$ 
after setting $i=j$, 
it is the character of a representation $\gamma$ 
of the Brauer algebra. 
} of a representation $\gamma$ of the Brauer algebra  
\begin{eqnarray}
\chi^{\gamma}_{\vec{r},ij}(b):=
\sum_{m=1}^{d_{\vec{r}}}\langle \gamma;\vec{r},m,i|b
| \gamma;\vec{r},m,i\rangle 
\end{eqnarray}
where 
$| \gamma;\vec{r},m,i\rangle$ is an orthonormal vector in 
the RHS of 
(\ref{Brauer_restriction_symmetric}).
We will not need the explicit form of $Q^{\gamma}_{\vec{r},ij}$ 
hereafter in this paper, 
but we will often use the following important properties 
\begin{eqnarray}
&&
h Q^{\gamma}_{\vec{r},ij} h^{-1}=Q^{\gamma}_{\vec{r},ij} \quad 
(h\in S_{n_1}\times S_{n_2}\times S_{n_3}),
\nonumber \\
&&
Q^{\gamma}_{\vec{r},ij}
Q^{\gamma^{\prime}}_{\vec{r}^{\prime},i^{\prime}j^{\prime}}
=
\delta^{\gamma \gamma^{\prime}}
\delta^{\vec{r}\vec{r}^{\prime}}
\delta_{ji^{\prime}}
Q^{\gamma}_{\vec{r},ij^{\prime}},
\nonumber \\
&& 
tr_{n_1+n_2,n_3}(Q^{\gamma}_{\vec{r},ij})=d_{\vec{r}} Dim\gamma \delta_{ij}. 
\label{projector_relation}
\end{eqnarray} 
$d_{\vec{r}}$ is the dimension of 
an irreducible representation $\vec{r}$ of 
$S_{n_1}\times S_{n_2}\times S_{n_3}$, and 
$Dim \gamma$ is the dimension of 
an irreducible representation $\gamma$ of 
$U(N)$. The third equation is a consequence of 
the Schur-Weyl duality (\ref{SWdual_Brauer}).

We will now show that two-point functions of the basis are diagonal. 
The two-point functions can be computed 
with the exact $N$-dependence
\begin{eqnarray}
\langle 
\gamma^{\prime},
\vec{r}^{\prime},i^{\prime}j^{\prime} |
\gamma,\vec{r},ij \rangle
&=&
\sum_{h \in H}
tr_{n_1+n_2,n_3}
(Q^{\gamma}_{\vec{r},ij}
h
Q^{\gamma^{\prime}}_{\vec{r}^{\prime},i^{\prime}j^{\prime}}{}^{\dagger}
h^{-1})
\nonumber \\
&=&
n_{1}!n_{2}!n_{3}!
tr_{n_1+n_2,n_3}
(Q^{\gamma}_{\vec{r},ij}
Q^{\gamma^{\prime}}_{\vec{r}^{\prime},j^{\prime}i^{\prime}}
)
\nonumber \\
&=&
n_{1}!n_{2}!n_{3}!
\delta_{\gamma \gamma^{\prime}}
\delta_{\vec{r} \vec{r}^{\prime}}
\delta_{ii^{\prime}}\delta_{jj^{\prime}}
d_{\vec{r}} Dim\gamma, 
\label{twopointBrauer}
\end{eqnarray}
where 
$H=S_{n_1}\times S_{n_2}\times S_{n_3}$. 
The $N$-dependence is contained in $Dim\gamma$. 
To get the first equality,  
we have exploited the fact that 
the Wick-contractions can be easily 
performed 
with the help of the symmetric group \cite{0111222}. 
The formula we have used is as follows,   
\begin{eqnarray}
\langle 
X^{i_1}_{j_1}\cdots X^{i_n}_{j_n}
X^{\dagger}{}^{k_1}_{l_1}\cdots X^{\dagger}{}^{k_n}_{l_n}
\rangle =
 \sum_{\sigma\in S_n}
\left(\delta^{i_1}_{l_{\sigma(1)}}\delta_{j_{\sigma^{-1}(1)}}^{k_{1}}\right) 
\left(\delta^{i_2}_{l_{\sigma(2)}}\delta_{j_{\sigma^{-1}(2)}}^{k_{2}}\right)  
\cdots 
\left(\delta^{i_n}_{l_{\sigma(n)}}\delta_{j_{\sigma^{-1}(n)}}^{k_{n}}\right) .
\end{eqnarray}
Because we have three matrices $X,Y,Z$, 
the Wick-contractions are given by the elements in $H$.
Likewise we have 
\begin{eqnarray}
\hspace{-0.3cm}
\langle 0 |B^{i_1}_{j_1}\cdots B^{i_n}_{j_n}
B^{\dagger}{}^{k_1}_{l_1}\cdots B^{\dagger}{}^{k_n}_{l_n}
|0\rangle =
 \sum_{\sigma\in S_n}
\left(\delta^{i_1}_{l_{\sigma(1)}}\delta_{j_{\sigma^{-1}(1)}}^{k_{1}}\right) 
\left(\delta^{i_2}_{l_{\sigma(2)}}\delta_{j_{\sigma^{-1}(2)}}^{k_{2}}\right)  
\cdots 
\left(\delta^{i_n}_{l_{\sigma(n)}}\delta_{j_{\sigma^{-1}(n)}}^{k_{n}}\right) .
\end{eqnarray}

\quad 

For an element $b$
in the Brauer algebra $B_N(n_1+n_2,n_3)$, we have the formula
\begin{eqnarray}
\frac{1}{n_1!n_2!n_3!}
\sum_{h \in H}
h b h^{-1}
=\sum_{\gamma,\vec{r},ij}
 \frac{1}{d_{\vec{r}} }\chi^{\gamma}_{\vec{r},ij}(b)
Q^{\gamma}_{\vec{r},ij}.
 \label{formula_centralelementH}
\end{eqnarray}
Both sides of (\ref{formula_centralelementH}) commute with any elements in $H$.
The formula leads to 
the following equation
\begin{eqnarray}
tr_{n_{1}+n_{2},n_3}(b X\otimes Y \otimes Z^T)
=
\sum_{\gamma,r,ij}
\frac{1}{d_r}
\chi^{\gamma}_{r,ij}
(b)
tr_{n_{1}+n_{2},n_3}(Q^{\gamma}_{r,ij} X\otimes Y 
\otimes Z^T). 
\label{completeness_equation}
\end{eqnarray}
Note that there are equivalence classes under 
conjugation of the group action of $H$. 
\footnote{
We have 
$tr_{n_{1}+n_{2},n_3}(b X\otimes Y \otimes Z^T)=
tr_{n_{1}+n_{2},n_3}(b^{\prime} X\otimes Y \otimes Z^T)$ 
if $b^{\prime}=hbh^{-1}$. Note also 
$\chi^{\gamma}_{r,ij}(b)=\chi^{\gamma}_{r,ij}(b^{\prime})$.
}
Taking into account that  
any gauge invariant operator built from $X$'s, $Y$'s, and $Z$'s
can be expressed in terms of an element of the Brauer algebra 
in the form of the left-hand side of the above equation, 
the above equation means the completeness of
the basis.

\quad 

When we have $N\times N $ matrices, 
we always have some relations between 
multi-traces. 
A simple example is $tr(\phi^3)=3/2 tr(\phi)tr(\phi^2)
-1/2 (tr\phi)^3$ for a $2\times 2$ matrix $\phi$. 
In general, $tr(\phi^p)$ ($p >N$) can be 
expressed by a linear combination of 
 $tr(\phi^q)$ ($q \le N$). 
An advantage of using representation bases 
is that finite $N$ relations are more manifest
because they are expressed by 
constraints on the Young diagrams. 
For the Brauer basis, we have constraints 
\begin{eqnarray}
&&
c_1(\gamma_+)+c_1(\gamma_-)\le N, 
\nonumber \\
&&
c_1(r_1)\le N, \quad  
c_1(r_2)\le N, \quad  
c_1(r_3)\le N ,
\label{cutoff_Brauer}
\end{eqnarray}
where 
$c_{1}(r)$ represents 
the first column length of a Young diagram $r$.

\quad 

Hereafter we will present  
concrete forms of the operator 
at two sectors 
in which $k$ takes the possible maximum value 
and the possible minimum value.  
This will enable us to guess the meaning of the integer $k$.  
The derivations being completely analogous to 
what we did in \cite{0709.2158,0807.3696,0911.4408},  
we will only show the final forms with leaving  
concrete derivations in 
appendix 
\ref{sec:specialsectors_Brauer}. 

\quad

In the $k=0$ sector, $\gamma_-$ and 
$r_3$ are identified, 
which we find from  
(\ref{multiplicity_brauerop}). 
The $k=0$ operators take the form  
\begin{eqnarray}
&&
tr_{n_1+n_2,n_3}(Q^{\gamma(k=0)}_{\vec{r},ij} 
X\otimes Y\otimes Z^{T}
)
\nonumber \\
&=&
Dim\gamma 
\frac{(n_1+n_2) !}{d_{\gamma_{+}}}
\frac{n_3 !}{d_{r_3}}
\frac{1}{N^{n}}
tr_{n}(\Omega_{n}^{-1}
P^{\gamma_{+}}_{(r_1,r_2),ij} p_{\gamma_-}
X^{\otimes n_1}\otimes Y^{\otimes n_2}\otimes Z^{\otimes n_3}
),
\label{k=0Brauer}
\end{eqnarray}
where $n:=n_1+n_2+n_3$. 
Note that $Z$'s are not transposed in the second line. 
Because $\Omega_{n}^{-1}/N^{n}$ 
starts from 1 in a $1/N$-expansion, 
the leading term with respect to $1/N$ is 
the product of 
a restricted Schur polynomial built from 
$P^{\gamma_{+}}_{(r_1,r_2),ij}$ 
and 
a Schur 
polynomial built from $p_{\gamma_-}$:
\begin{eqnarray}
&&
\frac{1}{N^{n}}
tr_{n}(\Omega_{n}^{-1}
P^{\gamma_{+}}_{(r_1,r_2),ij} p_{\gamma_-}
X^{\otimes n_1}\otimes Y^{\otimes n_2}\otimes Z^{\otimes n_3}
)
\nonumber \\
&=& 
tr_{n_1+n_2}(
P^{\gamma_{+}}_{(r_1,r_2),ij} 
 X^{\otimes n_1}\otimes Y^{\otimes n_2}
)
tr_{n_3}(
p_{\gamma_-}Z^{\otimes n_3})+\cdots .
\label{leading_k=0}
\end{eqnarray}
This implies that 
the Brauer basis is suitable for 
organising a composite of two operators. 

Because the above expression (\ref{k=0Brauer}) is written 
in terms of the tensor product appeared in 
the construction of the restricted Schur basis, 
we can easily express 
(\ref{k=0Brauer}) as a linear combination of 
the restricted Schur operators
\begin{eqnarray}
&&
\hspace{-1cm}
tr_{n_1+n_2,n_3}(Q^{\gamma(k=0)}_{\vec{r},ij} 
X\otimes Y\otimes Z^{T}
)=
\nonumber \\
&&
\hspace{-1cm}
Dim\gamma 
\frac{(n_1+n_2) !n_3 !}{n!}
\frac{1}{d_{\gamma_+}d_{r_{3}}}
\sum_{S\vdash n,k,l}
\frac{1}{d_{\vec{r}}}
\chi_{\vec{r},kl}^S
\left(P^{\gamma_{+}}_{(r_1,r_2),ij}
\right)
\frac{d_S}{DimS}
tr_{n}(
P^S_{\vec{r},kl} 
X^{\otimes n_1}\otimes Y^{\otimes n_2}\otimes Z^{\otimes n_3}
). 
\end{eqnarray}
The two operators have 
the same representation $\vec{r}$.

\quad 

We next look at the case in which 
$k$ takes the maximum possible value. 
Suppose $n_1+n_2=n_3$, for simplicity. 
When $k$ takes 
the maximum value 
$k=n_1+n_2=n_3$, 
$\gamma$ is $(\emptyset,\emptyset)$, and 
the multiplicity labels $i$,$j$ 
run over $1,\cdots,g(r_1,r_2;r_3)$. 
The explicit form in this sector is given by 
\begin{eqnarray}
&&
tr_{n_1+n_2,n_3}(
Q^{\gamma (k=n_1+n_2=n_3)}_{(r_1,r_2,r_3),ij}
X^{\otimes n_1}\otimes Y^{\otimes n_2} \otimes Z^{T\otimes n_3}
)
\nonumber \\
&=&
\frac{d_{r_3}}{Dim r_3 }
tr_{n_1+n_2}(
 P^{r_3}_{(r_1,r_2),ij}
(ZX)^{\otimes n_1} \otimes (ZY)^{\otimes n_2} 
).
\label{maximal_k_operator}
\end{eqnarray}
This is nothing but 
a restricted Schur operator built from  
two matrices $ZX$ and $ZY$. Hence 
the operator (\ref{maximal_k_operator}) 
looks describing open string excitations 
on giant gravitons 
where their angular momenta 
are excited along some directions. 
This sector might be helpful to 
describe a configuration involving 
such combined matrices.

\quad 

We have seen the two sectors of the Brauer basis, 
where $k$ takes the maximum possible value 
and the minimum possible value. 
From the two examples, 
we may be led to an interpretation of $k$ 
that it corresponds a degree of freedom describing  
the mixing between 
the $X$-$Y$ sector 
and the $Z$ sector. 
In \cite{1109.2585}
the integer $k$ was given a meaning  
in the context of quarter BPS bubbling geometries, 
where  
it was conjectured that the mixing between the two angular directions of 
the geometries is labelled by $k$.

\quad 

Let us now comment on 
including anti-holomorphic matrices. 
Originally we guessed that 
the Brauer basis 
would be a good basis for  
describing a system that contains
giant gravitons and anti-giant gravitons \cite{0709.2158}. 
With the same spirit, describing 
a more general system of giant gravitons 
and anti-giant gravitons 
is naturally given by 
considering the Brauer algebra 
$B_{N}(n_1+n_2+n_3,\bar{n}_1+\bar{n}_2+\bar{n}_3)$
\begin{eqnarray}
tr_{n_1+n_2+n_3,\bar{n}_1+\bar{n}_2+\bar{n}_3}(
Q^{\gamma}_{\vec{r},ij}
X^{\otimes n_1}\otimes 
Y^{\otimes n_2}\otimes
Z^{\otimes n_3} \otimes 
X^{*\otimes \bar{n}_1}\otimes Y^{*\otimes \bar{n}_2}\otimes
Z^{*\otimes  \bar{n}_3}
),
\end{eqnarray}
where $\vec{r}=(s_1,s_2,s_3,t_1,t_2,t_3)$ is an irreducible 
representation of $\mathbb{C}[S_{n_1}\times 
S_{n_2}\times S_{n_3}\times S_{\bar{n}_1}
\times S_{\bar{n}_2}\times S_{\bar{n}_3}]$. 
The $\gamma$ is an irreducible representation of the Brauer algebra; 
$\gamma_{+}\vdash (n_1+n_2+n_3-k)$, 
$\gamma_{-}\vdash (\bar{n}_1+\bar{n}_2+\bar{n}_3-k)$, 
$0\le k \le min(n_1+n_2+n_3,\bar{n}_1+\bar{n}_2+\bar{n}_3)$. 
The leading term of the $k=0$ operators is indeed a product of two operators 
corresponding to the holomorphic sector and the anti-holomorphic sector, which 
can be written up to a numerical factor as
\begin{eqnarray}
tr_{n_1+n_2+n_3}(
P^{\gamma_+}_{\vec{s},kl}
X^{\otimes n_1}\otimes 
Y^{\otimes n_2}\otimes
Z^{\otimes n_3} 
)
 tr_{\bar{n}_1+\bar{n}_2+\bar{n}_3}(P^{\gamma_-}_{\vec{t},pq}
X^{\dagger \otimes \bar{n}_1}\otimes Y^{\dagger\otimes \bar{n}_2}\otimes
Z^{\dagger\otimes  \bar{n}_3}
).
\end{eqnarray}


We also have an option to utilise the symmetric group 
$S_{n_1+n_2+n_3+\bar{n}_1+\bar{n}_2+\bar{n}_3}$ 
\cite{0801.2061} 
\begin{eqnarray}
tr_{n_1+n_2+n_3,\bar{n}_1+\bar{n}_2+\bar{n}_3}(
P^{R}_{\vec{r},ij}
X^{\otimes n_1}\otimes 
Y^{\otimes n_2}\otimes
Z^{\otimes n_3} \otimes 
X^{\dagger\otimes \bar{n}_1}\otimes Y^{\dagger \otimes \bar{n}_2}\otimes
Z^{\dagger\otimes  \bar{n}_3}
).
\end{eqnarray}
It is also possible to use another Brauer algebra such as 
$B_{N}(n_1+n_2+n_3+\bar{n}_1, \bar{n}_2+\bar{n}_3)$
\begin{eqnarray}
tr_{n_1+n_2+n_3+\bar{n}_1,\bar{n}_2+\bar{n}_3}(
Q^{\gamma^{\prime}}_{\vec{r},ij}
X^{\otimes n_1}\otimes 
Y^{\otimes n_2}\otimes
Z^{\otimes n_3} \otimes 
X^{\dagger \otimes \bar{n}_1}\otimes Y^{*\otimes \bar{n}_2}\otimes
Z^{*\otimes  \bar{n}_3}
),
\end{eqnarray}
where $\gamma^{\prime}_{+}\vdash (n_1+n_2+n_3+\bar{n}_1-k)$ and 
$\gamma^{\prime}_{-}\vdash (\bar{n}_2+\bar{n}_3-k)$, 
$0\le k \le min(n_1+n_2+n_3+\bar{n}_1,\bar{n}_2+\bar{n}_3)$. 
Another kind of basis of non-holomorphic operators
was given in
\cite{0910.2170} 
by combining a Brauer algebra and 
the method in \cite{0711.0176,0806.1911}.


\section{Multi-point correlation functions}
\label{sec:correlation_functions}

In this section, we will be concerned with multi-point
correlation functions 
of the Brauer operators. 
Multi-point functions can be computed 
efficiently 
by a product rule \cite{0805.3025}, which allows us to carry out the computation of
multi-point functions 
in terms of the two-point function we obtained in 
(\ref{twopointBrauer}). 

We first consider the 
holomorphic operators. 
The product rule of this case is as follows
\begin{eqnarray}
&& 
tr_{n_{X1}+n_{Y1},n_{Z1}}(Q^{\gamma_1}_{\vec{r}_1,i_1j_1}
X^{\otimes n_{X1}}\otimes Y^{\otimes n_{Y1}}
\otimes Z^{T \otimes n_{Z1}})
\nonumber \\
&&
\times 
tr_{n_{X2}+n_{Y2},n_{Z2}}(Q^{\gamma_2}_{\vec{r}_2,i_2j_2}
X^{\otimes n_{X2}}\otimes Y^{\otimes n_{Y2}}
\otimes Z^{T \otimes n_{Z2}})
\nonumber \\
&=&
tr_{n_X+n_Y,n_Z}(
(Q^{\gamma_1}_{\vec{r}_1,i_1j_1}\circ 
Q^{\gamma_2}_{\vec{r}_2,i_2j_2})
X^{\otimes n_X}\otimes Y^{\otimes n_Y}
\otimes Z^{T \otimes n_Z})
\nonumber \\
&=&
\sum_{\gamma,\vec{r},ij} 
\frac{1}{d_{\vec{r}}}
\chi^{\gamma}_{\vec{r},ij}
(Q^{\gamma_1}_{\vec{r}_1,i_1j_1}\circ 
Q^{\gamma_2}_{\vec{r}_2,i_2j_2})
tr_{n_X+n_Y,n_Z}(
Q^{\gamma}_{\vec{r},ij}
X^{\otimes n_X}\otimes Y^{\otimes n_Y}
\otimes Z^{T \otimes n_Z}), 
\label{productrule_Brauer}
\end{eqnarray}
where 
$n_{X}=n_{X1}+n_{X2}$, 
$n_{Y}=n_{Y1}+n_{Y2}$, 
$n_{Z}=n_{Z1}+n_{Z2}$.
To get the last equality we have exploited 
(\ref{completeness_equation}). 
Thanks to the product rule, 
we immediately obtain the following expression for 
three-point functions
\begin{eqnarray}
\langle 
O^{\gamma_1}_{\vec{r}_1,i_1j_1}
O^{\gamma_2}_{\vec{r}_2,i_2j_2}
O^{\gamma}_{\vec{r},ij}{}^{\dagger}
 \rangle 
=
n_X!n_Y!n_Z!
Dim \gamma \hspace{0.1cm}
\chi^{\gamma}_{\vec{r},ij}
(Q^{\gamma_1}_{\vec{r}_1,i_1j_1}\circ 
Q^{\gamma_2}_{\vec{r}_2,i_2j_2}). 
\label{threept_function_brauer}
\end{eqnarray}
Note that the expectation value is evaluated by the matrix model. 
For reference, we write  
the two-point function in a similar form  
\begin{eqnarray}
\langle 
O^{\gamma_1}_{\vec{r}_1,i_1j_1}
O^{\gamma_2}_{\vec{r}_2,i_2j_2}{}^{\dagger}
\rangle 
=
n_X!n_Y!n_Z!
Dim \gamma_2 \hspace{0.1cm}
\chi^{\gamma_2}_{\vec{r}_2,i_2j_2}
( 
Q^{\gamma_1}_{\vec{r}_1,i_1j_1}). 
\end{eqnarray}
The result can also be extended to multi-point functions
\begin{eqnarray}
&&
\langle 
O^{\gamma_1}_{\vec{r}_1,i_1,j_1}
\cdots
O^{\gamma_{s}}_{\vec{r}_{s},i_{s}j_{s}}
O^{\gamma_{s+1}}_{\vec{r}_{s+1},i_{s+1}j_{s+1}}{}^{\dagger}
\cdots
O^{\gamma_{t}}_{\vec{r}_{t},i_{t}j_{t}}{}^{\dagger}
 \rangle 
 \nonumber \\
 &&
 \hspace{-0.8cm}
=
n_X!n_Y!n_Z!
\sum_{\gamma,\vec{r},i,j}
Dim \gamma \hspace{0.1cm}
\frac{1}{d_{\vec{r}}}
\chi^{\gamma}_{\vec{r},ij}
(Q^{\gamma_1}_{\vec{r}_1,i_1j_1}\circ 
\cdots\circ 
Q^{\gamma_{s}}_{\vec{r}_{s},i_{s}j_{s}}
)
\chi^{\gamma}_{\vec{r},ji}
(Q^{\gamma_{s+1}}_{\vec{r}_{s+1},j_{s+1}i_{s+1}}\circ 
\cdots\circ 
Q^{\gamma_{t}}_{\vec{r}_{t},j_{t}i_{t}}
),
\end{eqnarray}
where $n_X=n_{X1}+\cdots+n_{Xs}=n_{X_{s+1}}+\cdots+n_{Xt}$, 
and $n_Y$ and $n_Z$ 
are also similarly defined. 
For comparison, we recall the result on the restricted 
Schur basis
(in our notation)
\cite{0805.3025}
\begin{eqnarray}
\langle 
O^{R_1}_{\vec{r}_1,i_1j_1}
O^{R_2}_{\vec{r}_2,i_2j_2}
O^{R}_{\vec{r},ij}{}^{\dagger}
 \rangle 
=
n_X!n_Y!n_Z!
Dim R \hspace{0.1cm}
\chi^{R}_{\vec{r},ij}
(P^{R_1}_{\vec{r}_1,i_1j_1}\circ 
P^{R_2}_{\vec{r}_2,i_2j_2}).
\label{rescticted_threept}
\end{eqnarray}
The character was concretely 
evaluated for some cases in \cite{0805.3025}. 

In this way, 
the evaluation of correlation functions 
has been replaced with the evaluation of 
the characters of symmetric groups or Brauer algebras. 

\quad 

We will now 
provide more concrete forms of 
the correlators 
for operators in 
the two specific sectors. 
Let us consider 
a correlator of the $k=0$ operators. 
Recall that the $k=0$ representations 
of the Brauer algebra
are factorised to be 
\begin{eqnarray}
V_{\gamma(k=0)}^{B_N(n_1+n_2,n_3)}
=V_{\gamma_+}^{\mathbb{C}[S_{n_1+n_2}]}\otimes 
V_{\gamma_-}^{\mathbb{C}[S_{n_3}]}. 
\end{eqnarray}
This means 
\begin{eqnarray}
\chi^{\gamma(k=0)}(b)=0
\label{k=0character}
\end{eqnarray}
for $b$ that is not an element in the subalgebra 
$\mathbb{C}[S_{n_1+n_2}]\times\mathbb{C}[ S_{n_3}]$.
This forces $k_1$, $k_2$ to be zero in order to get a non-zero correlator.
This property allows us to rewrite 
the character as a form where  
$\gamma_+$-part and the $\gamma_-$-part are factored
\begin{eqnarray}
\chi^{\gamma(k=0)}_{\vec{r},ij}
(Q^{\gamma_1(k=0)}_{\vec{r}_1,i_1j_1}\circ 
Q^{\gamma_2(k=0)}_{\vec{r}_2,i_2j_2})
=
\chi^{\gamma^{+}}_{(r_{1},r_2),ij}
(P^{\gamma_{1+}}_{(r_{11},r_{12}),i_1j_1}\circ 
P^{\gamma_{2+}}_{(r_{21},r_{22}),i_2j_2}
)
d_{r_{13}}d_{r_{23}}
g(r_{13},r_{23};r_3),
\label{factorised}
\end{eqnarray}
where 
$\vec{r}_1=(r_{11},r_{12},r_{13})$, 
$\vec{r}_2=(r_{21},r_{22},r_{23})$. 
The first factor in the right-hand side is the factor appearing 
in the three-point function of 
the restricted Schur-polynomials built from 
two matrices $X,Y$ 
(see (\ref{rescticted_threept})), while 
the second factor is a factor appearing 
in the three-point function of the Schur-polynomials 
built from $Z$ (see \cite{0111222}). 
In the next section, we will give a systematic study 
of a condition 
a class of correlators are factorised.

When all $k$'s have the possible maximum values
with conditions $n_X+n_Y=n_Z$, 
$n_{X1}+n_{Y1}=n_{Z1}$, and $n_{X2}+n_{Y2}=n_{Z2}$, 
that is,  
all operators are expressed by the form
(\ref{maximal_k_operator}), 
we have 
\begin{eqnarray}
&&
\langle 
O^{\gamma_1(k_1=n_{Z1})}_{\vec{r}_1,i_1j_1}
O^{\gamma_2 (k_2=n_{Z2})}_{\vec{r}_2,i_2j_2}
O^{\gamma (k=n_Z)}_{\vec{r},ij}{}^{\dagger}
 \rangle 
 \nonumber \\
&=&
n_X!n_Y!n_Z!
\frac{Dimr_3}{Dim r_1 Dimr_2}d_{r_1}d_{r_2}
\chi^{r_3}_{(r_1,r_{2}),ij}(P^{r_{13}}_{(r_{11},r_{12}),i_1j_1}\circ 
P^{r_{23}}_{(r_{21},r_{22}),i_2j_2}). 
\end{eqnarray}
It is non-zero if $k_1+k_2=k$, which comes from R-charge conservation. 
From (\ref{rescticted_threept}),  
the above character factor implies that 
the correlator is equivalent to 
a correlator of the restricted Schur operators built from two matrices.

\qquad

From now onwards we include non-holomorphic operators. 
In order to illustrate how to compute such 
correlation functions, 
we will consider the following as a simple example
\begin{eqnarray}
\langle 
:tr_{m_1,n_1}(Q^{\gamma_1}_{\vec{r_1},i_1j_1}
X^{\otimes m_1}\otimes X^{* \otimes n_1}):
:tr_{m_2,n_2}(Q^{\gamma_2}_{\vec{r_2},i_2j_2}
X^{\otimes m_2}\otimes X^{* \otimes n_2}):
:tr_{m,n}(Q^{\gamma}_{\vec{r},ji}
X^{\dagger \otimes m}\otimes X^{T \otimes n}):
 \rangle .
 \label{nonholo_correlator}
\end{eqnarray}
From R-charge conservation, 
it is zero unless  
$m_1+m_2+n=n_1+n_2+m$ is satisfied. 

If we have 
$m=m_{1}+m_{2}$ and $n=n_{1}+n_{2}$, 
$X^{\dagger}$'s (or $X^T$'s) in the third operator 
have to be contracted with 
$X$'s (or $X^{\ast}$'s) in the first one  
and $X$'s (or $X^{\ast}$'s) in the second one, 
and   
we do not have any contractions between 
the first operator and the second operator. 
We are then allowed to  
use the product rule 
(\ref{productrule_Brauer}) to  
get a similar result to (\ref{threept_function_brauer}). 
But if $m=m_{1}+m_{2}$, $n=n_{1}+n_{2}$ are not satisfied, 
Wick-contractions between the first operator 
and the second operator 
are missed if the product rule (\ref{productrule_Brauer}) is 
naively applied to this case.

The product rule for the case under consideration is given by
\begin{eqnarray}
&&
:tr_{m_1,n_1}(Q^{\gamma_1}_{\vec{r_1},i_1j_1}
X^{\otimes m_1}\otimes X^{* \otimes n_1}):
:tr_{m_2,n_2}(Q^{\gamma_2}_{\vec{r_2},i_2j_2}
X^{\otimes m_2}\otimes X^{* \otimes n_2}):
\nonumber \\
&=&
e^{
tr(\partial_X \partial_{Y^{\dagger}})
+tr(\partial_Y \partial_{X^{\dagger}})
}
:tr_{m_1,n_1}(Q^{\gamma_1}_{\vec{r_1},i_1j_1}
X^{\otimes m_1}\otimes X^{* \otimes n_1})
tr_{m_2,n_2}(Q^{\gamma_2}_{\vec{r_2},i_2j_2}
Y^{\otimes m_2}\otimes Y^{* \otimes n_2}):
_{\substack{[Y=X] \\ [Y^{\dagger}=X^{\dagger}]}}. 
\nonumber 
\end{eqnarray}
The symbol 
$[Y=X], [Y^{\dagger}=X^{\dagger}]$
appearing at the lower right of the equation means  
we set $Y=X, Y^{\dagger}=X^{\dagger}$
after performing the exponential operation. 
This is Wick's theorem. 

In order to see
the effect of 
the exponential factor, 
let us consider 
\begin{eqnarray}
&&
tr(\partial_X \partial_{Y^{\dagger}})
:tr_{m_1,n_1}(Q^{\gamma_1}_{\vec{r_1},i_1j_1}
X^{\otimes m_1}\otimes X^{* \otimes n_1})
tr_{m_2,n_2}(Q^{\gamma_2}_{\vec{r_2},i_2j_2}
Y^{\otimes m_2}\otimes Y^{* \otimes n_2}):
_{[Y=X, Y^{\dagger}=X^{\dagger}]}
\nonumber \\
&=&
\sum_{1\le a \le m_1, n_1+1\le b \le n_1+n_2}
:tr_{m_1+m_2,n_1+n_2}(
C_{ab} (Q^{\gamma_1}_{\vec{r_1},i_1j_1}
\circ Q^{\gamma_2}_{\vec{r_2},i_2j_2})
\nonumber \\
&& \hspace{0.4cm}
\times 
X^{\otimes a-1}\otimes 1\otimes X^{m_1+m_2-a} \otimes 
X^{* \otimes b-1}\otimes 1\otimes X^{* \otimes n_1+n_2-b}
):,
\label{contribution1}
\end{eqnarray}
where 
$C_{ab}$ is the contraction operator acting on 
the $a$-th $X$ and the $b$-th $X^{\ast}$ 
\begin{eqnarray}
tr(\partial_X \partial_{X^{\dagger}})
X^{i}_{j}X^{\ast}{}^{l}_k=\delta^{il}\delta_{jk}=(C_{ab})_{jk}^{il}. 
\end{eqnarray}
The action of $C_{ab}$ 
($1\le a \le m_1, n_1+1\le b \le n_1+n_2$) 
on 
$(Q^{\gamma_1}_{\vec{r}_1,i_1j_1}\circ 
Q^{\gamma_2}_{\vec{r}_2,i_2j_2})$ is explicitly written down as 
\begin{eqnarray}
&
C_{ab}&:(Q^{\gamma_1}_{\vec{r}_1,i_1j_1}\circ 
Q^{\gamma_2}_{\vec{r}_2,i_2j_2})
^{i_1\cdots \underline{i_{a}} \cdots 
i_{m_1} i_{m_1+1} \cdots i_{m_1+n_1};
k_1\cdots k_{m_2} k_{m_2+1} \cdots \underline{k_{b}}\cdots k_{m_2+n_2}
}
_{j_1\cdots j_{a} \cdots 
j_{m_1} j_{m_1+1} \cdots j_{m_1+n_1};
l_1\cdots l_{m_2} l_{m_2+1} \cdots l_{b}\cdots l_{m_2+n_2}
}
\nonumber \\
&\rightarrow &
\delta_{i_a k_b} 
\sum_s
( Q^{\gamma_1}_{\vec{r}_1,i_1j_1}\circ 
Q^{\gamma_2}_{\vec{r}_2,i_2j_2})
^{i_1\cdots \underline{s} \cdots 
i_{m_1} i_{m_1+1} \cdots i_{m_1+n_1};
k_1\cdots k_{m_2} k_{m_2+1} \cdots \underline{s}\cdots k_{m_2+n_2}
}
_{j_1\cdots j_{a} \cdots 
j_{m_1} j_{m_1+1} \cdots j_{m_1+n_1};
l_1\cdots l_{m_2} l_{m_2+1} \cdots l_{b}\cdots l_{m_2+n_2}
}.
\end{eqnarray}
Similarly we have 
\begin{eqnarray}
&&
tr(\partial_Y \partial_{X^{\dagger}})
:tr_{m_1,n_1}(Q^{\gamma_1}_{\vec{r_1},i_1j_1}
X^{\otimes m_1}\otimes X^{* \otimes n_1})
tr_{m_2,n_2}(Q^{\gamma_2}_{\vec{r_2},i_2j_2}
Y^{\otimes m_2}\otimes Y^{* \otimes n_2}):_{[Y=X, Y^{\dagger}=X^{\dagger}]}
\nonumber \\
&=&
\sum_{m_1+1\le a \le m_1+m_2,1\le b \le n_1}
:tr_{m_1+m_2,n_1+n_2}(
C_{ab} (Q^{\gamma_1}_{\vec{r_1},i_1j_1}
\circ Q^{\gamma_2}_{\vec{r_2},i_2j_2})
\nonumber \\
&& \hspace{0.2cm}
\times 
X^{\otimes a-1}\otimes 1\otimes X^{m_1+m_2-a} \otimes 
X^{* \otimes b-1}\otimes 1\otimes X^{* \otimes n_1+n_2-b}
):.
\label{contribution2}
\end{eqnarray}
The above two terms 
(\ref{contribution1}) and (\ref{contribution2})
can be combined to give 
\begin{eqnarray}
\sum_{\gamma ,\vec{r},i,j} 
\frac{1}{d_{\vec{r}}}
\chi^{\gamma}_{\vec{r},ij}
(D_{(1)}(Q^{\gamma_1}_{\vec{r}_1,i_1j_1}\circ 
Q^{\gamma_2}_{\vec{r}_2,i_2j_2}))
:tr_{m_1+m_2-1,n_1+n_2-1}(Q^{\gamma}_{\vec{r},ij}
X^{\otimes m_1+m_2-1}\otimes X^{* \otimes n_1+n_2-1}):,
\label{non-holo_productrule_C}
\end{eqnarray}
where $\vec{r}$ and $\gamma$ run over irreducible representations of 
$S_{m_1+m_2-1}\times S_{n_1+n_2-1}$ and 
irreducible representations of 
$B_{N}(m_1+m_2-1,n_1+n_2-1)$ respectively. 
To write down the above form, 
we have introduced 
an operation $D_{(1)}$
\begin{eqnarray}
D_{(1)}=\sum_{1\le a\le m_1 ,n_1+1 \le b\le n_1+n_2}D_{ab}
+
\sum_{m_1+1\le a\le m_1+m_2 ,1\le b\le n_1}D_{ab}, 
\end{eqnarray}
where 
\begin{eqnarray}
&
D_{ab}&:(Q^{\gamma_1}_{\vec{r}_1,i_1j_1}\circ 
Q^{\gamma_2}_{\vec{r}_2,i_2j_2})
^{i_1\cdots i_{a} \cdots 
i_{m_1} i_{m_1+1} \cdots i_{m_1+n_1};
k_1\cdots k_{m_2} k_{m_2+1} \cdots k_{b}\cdots k_{m_2+n_2}
}
_{j_1\cdots j_{a} \cdots 
j_{m_1} j_{m_1+1} \cdots j_{m_1+n_1};
l_1\cdots l_{m_2} l_{m_2+1} \cdots l_{b}\cdots l_{m_2+n_2}
}
\nonumber \\
&\rightarrow &
\sum_{s,t}
( Q^{\gamma_1}_{\vec{r}_1,i_1j_1}\circ 
Q^{\gamma_2}_{\vec{r}_2,i_2j_2})
^{i_1\cdots s \cdots 
i_{m_1} i_{m_1+1} \cdots i_{m_1+n_1};
k_1\cdots k_{m_2} k_{m_2+1} \cdots s\cdots k_{m_2+n_2}
}
_{j_1\cdots t \cdots 
j_{m_1} j_{m_1+1} \cdots j_{m_1+n_1};
l_1\cdots l_{m_2} l_{m_2+1} \cdots t \cdots l_{m_2+n_2}
},
\end{eqnarray}
for $1\le a\le m_1$, $n_1+1 \le b\le n_1+n_2$. 
Note that the operation $D_{ab}$ is not a linear map on 
$V^{\otimes m_1+m_2}\otimes \bar{V}^{\otimes n_1+n_2}$,  
while $C_{ab}$ is a linear map on it.

Thus the contribution of 
the term
(\ref{non-holo_productrule_C})
in the correlation function  
(\ref{nonholo_correlator}) yields
\begin{eqnarray}
\delta_{m,m_1+m_2-1}\delta_{n,n_1+n_2-1}
m!n!Dim\gamma
\hspace{0.1cm}
 \chi^{\gamma}_{\vec{r},ij}
\left(D_{(1)}(Q^{\gamma_1}_{\vec{r}_1,i_1j_1}\circ 
Q^{\gamma_2}_{\vec{r}_2,i_2j_2})\right).
\end{eqnarray}
Defining 
\begin{eqnarray}
D_{(s)}&=&
\sum_{p=0}^s
\frac{1}{(s-p)!p!}
\left(\sum_{1\le a_1 \neq \cdots \neq a_p \le m_1}
\sum_{n_1+1\le b_1\neq \cdots\neq b_p \le n_1+n_2}
D_{a_1b_1}\cdots D_{a_p b_p}
\right)
\nonumber \\
&&
\times 
\left(
\sum_{m_1+1\le a_1 \neq \cdots \neq a_{s-p} \le m_1+m_2}
\sum_{1\le b_1\neq \cdots\neq b_{s-p} \le n_1}
D_{a_1b_1}\cdots D_{a_{s-p} b_{s-p}}
\right),
\end{eqnarray}
the correlation function 
(\ref{nonholo_correlator}) can be computed as 
\begin{eqnarray}
&&
\langle 
:tr_{m_1,n_1}(Q^{\gamma_1}_{\vec{r_1},i_1j_1}
X^{\otimes m_1}\otimes X^{* \otimes n_1}):
:tr_{m_2,n_2}(Q^{\gamma_2}_{\vec{r_2},i_2j_2}
X^{\otimes m_2}\otimes X^{* \otimes n_2}):
:tr_{m,n}(Q^{\gamma}_{\vec{r},ij}
X^{*\otimes m}\otimes X^{ \otimes n}):
 \rangle 
\nonumber \\
&=&
m!n!Dim\gamma
\hspace{0.1cm}
 \sum_{0\le s \le m_1+m_2-m}
 \delta_{m,m_1+m_2-s}\delta_{n,n_1+n_2-s}
 \chi^{\gamma}_{\vec{r},ij}
\left(D_{(s)}(Q^{\gamma_1}_{\vec{r}_1,i_1j_1}\circ 
Q^{\gamma_2}_{\vec{r}_2,i_2j_2})\right).
\end{eqnarray}


\section{Correlation functions of the BPS operators}
\label{correlator_BPS}

In this section, 
we shall turn our interest to 
a class of 
operators which 
are labelled by $\gamma$ alone. 
Namely the operator we will consider is
\begin{eqnarray}
O^{\gamma}=
tr_{n_1+n_2,n_3}(P^{\gamma}X^{\otimes n_1}\otimes Y^{\otimes n_2}\otimes 
Z^{T\otimes n_3}),
\end{eqnarray}
where $\gamma=(\gamma_+,\gamma_-)\vdash (n_1+n_2-k,n_3-k)$ 
($0\le k\le min(n_1+n_2,n_3)$). 
The operator 
$P^{\gamma}=\sum_{\vec{r},i}Q^{\gamma}_{\vec{r},ii}$ 
is the projection operator for
an irreducible representation $\gamma$. 
This class can attract a special interest because 
they are annihilated by the one-loop dilatation operator \cite{1002.2424}. 
An essential part of
computations 
can be captured by the two-matrix case, and hence 
we will study the case $n_2=0$ (and rename $Z$ to $Y$, $n_3$ to $n_2$).

From the previous section, for three-point functions 
we have 
\begin{eqnarray}
&& \langle 
O^{\gamma_1}
O^{\gamma_2}
O^{\gamma}{}^{\dagger}
 \rangle 
\nonumber \\
&=&
\langle 
tr_{m_1+m_2,n_1+n_2}((
P^{\gamma_1}
\circ 
P^{\gamma_2}
)
X^{\otimes m_1+m_2}\otimes Y^{T\otimes n_1+n_2}
)
tr_{m,n}(
P^{\gamma}
X^{\dagger \otimes m}\otimes Y^{\ast\otimes n}
)
 \rangle 
 \nonumber \\
&=&
m!n! \delta_{m_1+m_2,m}\delta_{n_1+n_2,n}
Dim\gamma\hspace{0.1cm} \chi^{\gamma}(P^{\gamma_1}
\circ 
P^{\gamma_2})
 \nonumber \\
&=&
m!n! \delta_{m_1+m_2,m}\delta_{n_1+n_2,n}
Dim\gamma\hspace{0.1cm}
d_{\gamma_1}d_{\gamma_2}
M_{\gamma_1,\gamma_2}^{\gamma} .
\label{3ptfunctionbps}
\end{eqnarray}
$d_{\gamma}$ is  
the dimension of $\gamma$ considered 
as 
an irreducible representation of the Brauer algebra. 
In the last line above we have used the branching rule 
for 
$B_{N}(m,n)\rightarrow B_{N}(m_1,n_1)\times B_{N}(m_2,n_2)$
\cite{Koike1989,Halverson1996}
\begin{eqnarray}
V^{\gamma}\cong \bigoplus_{\gamma_1,\gamma_2} M_{\gamma_1,\gamma_2}^{\gamma}
V^{\gamma_1}\otimes V^{\gamma_2}, 
\end{eqnarray}
where $M_{\gamma_1,\gamma_2}^{\gamma}$ is the multiplicity that  
counts 
the number of times $\gamma$ appears in the direct product of 
$\gamma_1$ and $\gamma_2$: 
\begin{eqnarray}
M_{\gamma_1,\gamma_2}^{\gamma}
=
\sum_{\rho,\zeta,\theta,\kappa}
\left(
\sum_{\delta}
g(\delta,\rho;\gamma_{1+})
g(\delta,\zeta;\gamma_{2-})
\right)
\left(
\sum_{\epsilon}
g(\epsilon,\theta;\gamma_{1-})
g(\epsilon,\kappa;\gamma_{2+})
\right)
g(\rho,\kappa;\gamma_{+})
g(\zeta,\theta;\gamma_{-}),
\label{Brauer_branching}
\end{eqnarray}
where 
$g(\delta,\rho;\gamma_{1+})$'s are the Littlewood-Richardson coefficients.
If $\gamma_{1-}=0$, $\gamma_{2-}=0$, and $\gamma_{-}=0$, 
we have 
\begin{eqnarray}
M_{(\gamma_{1+},\emptyset),(\gamma_{2+},\emptyset)}^{
(\gamma_{+},\emptyset)}
=
g(\gamma_{1+},\gamma_{2+};\gamma_{+}),
\end{eqnarray}
as expected.

Up to the normalisation factor, the three-point function 
is equivalent to 
the multiplicity associated 
with the restriction. 
Therefore
information about the relevant physics 
is fully contained in the multiplicity and the normalisation. 
We shall try to extract physics by examining the multiplicity 
for several cases. 
In particular, 
one of our concerns is to know 
if there are relations among 
the integers $k$, $k_1$, $k_2$ 
for non-zero correlators.  
In order to manifest the value of $k$
we will often write 
representations like $\gamma=(\gamma_+,\gamma_-,k)$ or $\gamma(k)$.


\quad 

Analysing the condition for the 
multiplicity to be non-zero, 
we obtain an inequality for $k,k_1,k_2$:
\begin{eqnarray}
k\ge k_1+k_2.
\label{embedding_gamma12}
\end{eqnarray}
We give the derivation 
in appendix \ref{ref:derivationofkk1k2}. 
The equality of (\ref{embedding_gamma12}) is $k=k_1+k_2$, where 
the multiplicity takes the following form
\begin{eqnarray}
M_{(\gamma_{1+},\gamma_{1-},k_1),(\gamma_{2+},\gamma_{2-},k_2)}
^{(\gamma_{+},\gamma_{-},k_1+k_2)}
=
g(\gamma_{1+},\gamma_{2+};\gamma_{+})
g(\gamma_{1-},\gamma_{2-};\gamma_{-}).
\label{equality_of_inequality}
\end{eqnarray}
Here 
the $\gamma_+$-sector and the $\gamma_-$-sector are completely decoupled.
We call the form {\bf factorised form}. 
In fact, 
$k=k_1+k_2$ is a necessary and sufficient condition for 
the correlator to be factorised. 
For our convenience, 
we introduce a quantity 
\begin{eqnarray}
\Delta:=k-(k_1+k_2), 
\label{Delta_deviation}
\end{eqnarray}
which cannot be negative due to (\ref{embedding_gamma12}). 
It measures 
the deviation of $k$ from $k_1+k_2$.  
Because $\Delta=0$ is the case the correlator takes a factorised form, 
we expect that it 
can be a good index to measure 
how far the correlator 
is from the factorised form. 
We will see 
if $\Delta$ is really a good index 
in some concrete situations.

When $\gamma$ takes the $k=0$ representation, 
$k_1$ and $k_2$ are forces to be at $k_1=0$ and $k_2=0$ 
for the multiplicity to be non-zero. 
The multiplicity takes the factorised form 
\begin{eqnarray}
&&
M_{\gamma_1(k=0),\gamma_2(k=0)}^{\gamma(k=0)}=
g(\gamma_{1+},\gamma_{2+};\gamma_+)
g(\gamma_{1-},\gamma_{2-};\gamma_-).
\label{k=0factorised3pt} 
\end{eqnarray}
Here the number of boxes in each Young diagram is equal to the R-charge. 
The factorised form looks like   
we have two copies of the 1/2 BPS sector.

\quad

Consider the case where 
$\gamma_1=(R,\emptyset)$, 
$\gamma_2=(\emptyset,S)$ 
($R\vdash m$, $S\vdash n$), i.e. 
the first operator is a Schur polynomial of $X$ 
while the second operator is another Schur polynomial 
of $Y$. 
The multiplicity becomes 
\begin{eqnarray}
M_{R,S}^{(\gamma_+,\gamma_-,k)}
=
\sum_{\delta \vdash k}
g(\delta,\gamma_+;R)
g(\delta,\gamma_-;S).
\label{multiplicity_Brauer_symmetric}
\end{eqnarray}
In this case 
$\gamma$ is allowed to take 
all possible values of 
$k$ 
to have non-zero transitions. 

We now try to give physical interpretations of the 
third operator. Suppose $m$ and $n$ are both ${\cal O}(N)$, and 
$R$ and $S$ are the symmetric representations or the anti-symmetric representations. 
These are just to have a concrete situation. 
The first operator and the second operator represent giant gravitons
expanding in the $S^5$ or in the $AdS_5$, but 
they have different angular directions, call $J_1$ and $J_2$.   
For $k=0$, we have non-zero transitions if and 
only if $\gamma_+=R$ and $\gamma_-=S$. 
The third operator can be naturally considered to be 
a giant graviton with two angular momenta $J_1$ and $J_2$ 
or a composite of the two giant gravitons. 
If $R$ and $S$ are totally anti-symmetric, 
a group theoretic prediction 
(coming from the first one in (\ref{cutoff_Brauer}))
is that  
the sum of the angular momenta should have cut-off at $N$. 
(This was called non-chiral stringy exclusion principle in \cite{0709.2158}.)
The correlator can also be non-zero 
for $k\ge 1$.  
For $k\ge 1$, we come across a new situation in which the size of the Young diagrams 
does not represent the R-charge. 
When $k$ is small, we give 
an interpretation that 
the third operator describes a system of a giant graviton whose size 
is determined by 
$\gamma_+$ and $\gamma_-$ and a closed string excitation 
determined by the $k$. 
Increasing the value of $k$ up to the possible maximum value, the third operator 
is well described in terms of combined matrices 
like in (\ref{maximal_k_operator}). 
It would be a giant graviton 
that is different from 
the giant graviton at $k=0$. 
In other words, a giant graviton whose size is determined by 
$k$ would emerge when $k$ is ${\cal O}(N)$. 
Because $\Delta=k$, increasing $k$ is increasing $\Delta$.

\quad 

We shall next discuss cases in which 
the second operator is a Schur polynomial. 
The simplest case is 
the restriction
$B_{N}(m,n)\rightarrow B_{N}(m-1,n)$, i.e. 
$\gamma_2=([1],0)$.
The operator labelled by 
$\gamma_2$ is just $trX$, representing a KK graviton 
with a unit of angular momentum. 
The relevant multiplicity is as follows
\begin{eqnarray}
&&
M_{(\gamma_{1+},\gamma_{1-},k_1),([1],\emptyset)}^{(\gamma_+,\gamma_-,k)}=
\sum_{\epsilon,\kappa}
g(\epsilon,\gamma_{-};\gamma_{1-})
g(\epsilon,\kappa;[1])
g(\gamma_{1+},\kappa;\gamma_{+})
.
\end{eqnarray}
There are two cases to get non-zero multiplicities, which are 
$(\epsilon,\kappa)=(\emptyset,[1])$ and 
$([1],\emptyset)$. For the first case, 
we have 
\begin{eqnarray}
M_{(R,S,k_1),([1],\emptyset)}
^{(R_+,S,k)}=1,
\end{eqnarray}
and zero otherwise. Here $S$ is a Young diagram with $n-k$ boxes 
and $R$ is a Young diagram with $m-1-k$ boxes.
$R_+$ is a Young diagram obtained by adding a box to the $R$. 
We have $k=k_1$ ($0\le k\le min(m-1,n)$). 
For the second case, 
we have 
\begin{eqnarray}
M_{(R,S_+,k_1),([1],\emptyset)}
^{(R,S,k)}=1, 
\end{eqnarray}
and zero otherwise. 
$R$ is a Young diagram with $m-k$ boxes 
and $S$ is a Young diagram with $n-k$ boxes, and we have  
$k=k_1+1$ ($1\le k\le min(m,n)$). 
In this way, we find 
that the branching rule does not allow $k$ and $k_1$ to take 
any possible values. They 
must be equal or 
be related by $k=k_1+1$. 
In terms of $\Delta$, 
$\Delta=0$ for the first case and $\Delta=1$ for the second case. 
\quad 

Generalising the above case, 
consider a more general case with $\gamma_{2-}=0$, i.e.  
$B_{N}(m,n)\rightarrow B_{N}(m_1,n)\times 
B_{N}(m_2,0)$ ($m=m_1+m_2$). We have
\begin{eqnarray}
&&
M_{\gamma_1,(T,\emptyset)}^{\gamma}=
\sum_{\epsilon,\kappa}
g(\epsilon,\gamma_{-};\gamma_{1-})
g(\epsilon,\kappa;T)
g(\gamma_{1+},\kappa;\gamma_{+}),
\label{branching_n2=0}
\end{eqnarray}
where $T\vdash m_2$. 
For the multiplicity to be non-zero, 
$k$ and $k_1$ cannot take any values, and  
the difference must satisfy the following 
\begin{eqnarray}
0 \le \Delta=k-k_1 \le m_2 ,
\end{eqnarray}
which we obtain by writing down consistency equations 
for the number of boxes of the Young diagrams 
in the LR coefficients. 
If we write $\gamma_1=(\alpha,\beta,k_1)$, where 
$\alpha\vdash (m_1-k_1)$, $\beta\vdash (n-k_1)$, 
non-zero multiplicities are obtained iff $\gamma$ is given by 
\begin{eqnarray}
\gamma=(\alpha_{+s},\beta_{-\Delta},k=k_1+\Delta) \quad (s=m_2-\Delta).
\end{eqnarray}
Here $\alpha_{+s}$ is a Young diagram obtained by the tensor product 
of the $\alpha$ and a Young diagram with $s$ boxes, 
and $\beta_{-\Delta}$ is a Young diagram that gives the 
$\beta$ when the tensor product with a Young diagram with $\Delta$ boxes 
is considered. 

When $\Delta=0$, 
all boxes in $T$ are added to $\alpha$. 
This is the case of $(\epsilon,\kappa)=(\emptyset,T)$ in (\ref{branching_n2=0}). 
The multiplicity is factorised 
\begin{eqnarray}
&&
M_{(\gamma_{1+},\gamma_{1-},k_1),(T,\emptyset)}^{(\gamma_+,\gamma_-,k=k_1)}=
g(\emptyset,\gamma_{-};\gamma_{1-})
g(\gamma_{1+},T;\gamma_{+}).
\end{eqnarray}
Shifting the value of $\Delta$ from zero, 
the correlator no longer takes a factorised form. 
Only $m_2-\Delta$ boxes are added to $\alpha$, and 
the remaining $\Delta$ boxes are added to $\beta_{-\Delta}$ to make $\beta$. 
When $\Delta=m_2$, all boxes in $T$ are added to $\beta_{-\Delta}$. 
Thus 
$\Delta$ is 
a good index to know how far the correlator is from the factorised form.

\quad

We next think about cases where $k$ takes the maximum possible value. 
Suppose $m=n$ just for simplicity. 
Because  
$\gamma=(\emptyset,\emptyset,k=m)$, 
the multiplicity takes the form 
\begin{eqnarray}
M_{\gamma_1,\gamma_2}^{\gamma}
&=&
\sum_{\delta,\epsilon}
g(\delta,\emptyset;\gamma_{1+})
g(\delta,\emptyset;\gamma_{2-})
g(\epsilon,\emptyset;\gamma_{1-})
g(\epsilon,\emptyset;\gamma_{2+})
\nonumber \\
&=&
g(\gamma_{1+},\emptyset;\gamma_{2-})
g(\gamma_{1-},\emptyset;\gamma_{2+}). 
\end{eqnarray}
This is non-zero 
if we have  $m_1-k_1=n_2-k_2$. 
We know that $k_1+k_2 =k=m_1+m_2=n_1+n_2$ 
is a necessary and sufficient condition for 
the correlator to be factorised. 
Considering 
$0\le k_1\le min(m_1,n_1)$ and $0\le k_2\le min(m_2,n_2)$, 
it is satisfied iff 
we have 
$m_1=n_1=k_1$ and 
$m_2=n_2=k_2$. 
Hence the case $(\gamma_{1+},\gamma_{1-})=(\emptyset,\emptyset)$ and 
$(\gamma_{2+},\gamma_{2-})=(\emptyset,\emptyset)$
is the only case the correlator is factorised.

When $k_2=m_2=n_2$ with 
arbitrary $\gamma_1$, $\gamma$, 
we have the non-zero multiplicities given by
\begin{eqnarray}
M_{(\gamma_{1+},\gamma_{1-},k_1),(\emptyset,\emptyset,k_2)}
^{(\gamma_{1+},\gamma_{1-},k)}
=1
\end{eqnarray}
with $m_1-k_1=m-k$. 
The correlator is already taking a factorised form. 
It is consistent because 
we always have 
\begin{eqnarray}
k=k_1+k_2,
\label{conserved_k}
\end{eqnarray}
which comes from 
R-charge conservation $m_1+m_2=m$.
It is interesting that (\ref{conserved_k}) is a reflection of 
R-charge conservation.

\quad 

Finally we present a more general  
correlation function 
\begin{eqnarray}
&&
\langle 
O^{\gamma_1}
\cdots
O^{\gamma_{s}}
O^{\gamma_{s+1}}{}^{\dagger}
\cdots
O^{\gamma_{s+t}}{}^{\dagger}
 \rangle 
 \nonumber \\
&=&
m!n!
\sum_{\gamma,\vec{r},ij}
Dim \gamma \hspace{0.1cm}
\frac{1}{d_{\vec{r}}}
\chi^{\gamma}_{\vec{r},ij}
(P^{\gamma_1}\circ 
\cdots\circ 
P^{\gamma_{s}}
)
\chi^{\gamma}_{\vec{r},ji}
(P^{\gamma_{s+1}}\circ 
\cdots\circ 
P^{\gamma_{s+t}}
). 
\end{eqnarray}
Here $\gamma$ is an irreducible representation of 
$B_N(m,n)$, and 
$\vec{r}$ is an irreducible representation of 
$\mathbb{C}[S_{m}\times S_n]$. We need 
$m=m_{1}+\cdots+m_{s}=m_{s+1}+\cdots+m_{s+t}$ and 
$n=n_{1}+\cdots+n_{s}=n_{s+1}+\cdots+n_{s+t}$ 
to obtain a non-zero result. 
When $t=1$, we get 
\begin{eqnarray}
\langle O^{\gamma_1} O^{\gamma_2} 
\cdots O^{\gamma_s} 
O^{\gamma}{}^{\dagger} \rangle 
&=&
m!n!
Dim \gamma \hspace{0.1cm}
\chi^{\gamma}
(P^{\gamma_1}\circ P^{\gamma_2}
\circ \cdots \circ P^{\gamma_s} )
\nonumber \\
&=&
m!n!
Dim \gamma \hspace{0.1cm}
d_{\gamma_1}d_{\gamma_{2}}
\cdots d_{\gamma_{s}}
M^{\gamma}_{\gamma_1,\gamma_2,\cdots, \gamma_{s}},
\label{multipt_bps}
\end{eqnarray}
where 
\begin{eqnarray}
M^{\gamma}_{\gamma_1,\gamma_2,\cdots, \gamma_{s}}:=
\sum_{r,q,.. ,r}
M^{\gamma}_{\gamma_1 p}
M^{p}_{\gamma_2 q}
\cdots 
M^{r}_{\gamma_{s-1} \gamma_{s}}. 
\end{eqnarray}
It is non-zero 
if the following inequality is satisfied
\begin{eqnarray}
k\ge k_1+k_2+\cdots +k_s,
\end{eqnarray}
which is an extension of (\ref{embedding_gamma12}). 
If the equality is the case, $k=k_1+k_2+\cdots +k_s$, 
the correlator takes the factorised form 
\begin{eqnarray}
\langle O^{\gamma_1} O^{\gamma_2} 
\cdots O^{\gamma_s} 
O^{\gamma}{}^{\dagger} \rangle 
=
m!n!
Dim \gamma \hspace{0.1cm}
g(\gamma_{1+},\gamma_{2+},\cdots;\gamma_+)
g(\gamma_{1-},\gamma_{2-},\cdots;\gamma_-). 
\end{eqnarray}
It is interesting that 
the matrix integral is decomposed into the two pieces.


\section{Discussions}
\label{sec:discussions}

In this paper, 
we have studied correlation functions of 
local gauge invariant operators in 
${\cal N}=4$ SYM at zero coupling by starting from the review of 
the symmetries of the free theory. 
In particular 
we have studied 
a basis that uses the Brauer algebra in more detail  
for the $so(6)$ scalar sector.
Our construction of bases followed the guideline that comes 
from the structure of commuting conserved charges. 

Inclusion of other fields is similarly managed. 
The Hamiltonian of the free theory on $R\times S^3$
can be written as a set of infinitely many 
harmonic oscillators \cite{0207067,0306054,0605163}.  
Suppose we are interested in an  
$s$-oscillator system. 
A representation basis can be labelled by  
two kinds of Young diagrams. 
One is a set of 
Young diagrams $(\alpha_1,\cdots,\alpha_s)$ 
corresponding to an irreducible representation of 
$S_{n_1}\times \cdots \times S_{n_s}$. 
This is related to the fact that 
the system has $s$ towers of conserved charges, 
corresponding to the set $S$ for the case of the $su(3)$ sector. 
The other is a Young diagram $R$ or 
a pair of Young diagrams $(\gamma_+,\gamma_-)$ corresponding to 
an irreducible representation of 
$S_{n_1+\cdots+n_s}$ or 
an irreducible representation of the Brauer algebra 
$B_{N}(n_+,n_-)$, where $n_++n_-=n_1+\cdots+n_s$. 
We may choose whichever basis we like in order to have 
an orthogonal basis, 
but it would be interesting to find that both have 
a common label 
corresponding to 
an irreducible representation of 
$S_{n_1}\times \cdots \times S_{n_s}$. 
This might suggest that  
we can furnish a universal physical meaning to the eigenvalues 
$C_p(\alpha_i)$ of 
the conserved charges 
in the context of 
a string theory or 
a higher-spin field theory on AdS.

In the latter part of this paper, we have computed
exact correlation functions of the Brauer basis at zero coupling. 
Multi-point correlation functions 
for a class of 1/4 BPS operators 
were written down by 
a branching rule of the Brauer algebra - see 
(\ref{3ptfunctionbps}) and (\ref{multipt_bps}).
We have found that  
the multi-point correlation functions 
take a factorised form 
in which the $\gamma_+$-sector and the $\gamma_-$-sector are completely decoupled 
if $k$'s satisfy a relation - the equality in 
(\ref{embedding_gamma12}). 
This might suggest a fascinating possibility that 
general correlation functions for the Brauer operators 
are well classified in terms of 
the integers. 
The meaning of $k$ was studied 
in the context of the correspondence 
between 1/4 BPS bubbling geometries and 
the 1/4 BPS Brauer operators in \cite{1109.2585}. 
It 
was naturally interpreted as 
the mixing between the two angular directions.  
It will be an interesting future direction to 
reinforce this interpretation from a study of 
correlation functions that contain large operators
using the techniques developed in \cite{0706.0216,0810.4041,0905.2273}.

We could see some similarities between 
the Brauer basis and the restricted Schur basis. 
They have a common set of conserved charges 
(which we denoted by $S$), and 
the Brauer basis contains the restricted Schur basis 
as a subset in the sense of 
(\ref{leading_k=0}) and (\ref{maximal_k_operator}). 
We are wondering if 
the Brauer operator admits 
a quantum non-planar integrability similar 
to the recent observations
in 
\cite{1004.1108,1012.3884,1101.5404,1106.2483,1108.2761,1111.1058,1111.6385,
1204.2153,1206.0813}.  See also \cite{1002.2099,1010.1683}
for the construction of BPS operators at weak coupling. 
It will be interesting to develop group theoretic methods 
for extracting universal features of non-planar theories.

\qquad 

\noindent 
{\bf Acknowledgements} 

I would like to thank Robert de Mello Koch and Sanjaye Ramgoolam
for reading 
the draft and making helpful comments.


\qquad 

\appendix 
\renewcommand{\theequation}
{\Alph{section}.\arabic{equation}}

\section{Representation theory of 
the Brauer algebra}
\setcounter{equation}{0}
\label{sec:irreps_Brauer}

In this appendix, we 
briefly summarise 
the representation theory of 
the Brauer algebra $B_N(m,n)$.  
See also the paper
\cite{0709.2158}, 
and references therein. 

The Brauer algebra, which we denote 
by $B_N(m,n)$,  
naturally appears 
in the decomposition of the following tensor product representation
\begin{eqnarray}
V^{\otimes m}
\otimes \bar{V}^{\otimes n}
&=&\bigoplus_{\gamma}
V_{\gamma}^{U(N)}
\otimes V_{\gamma}^{B_N(m,n)},
\end{eqnarray} 
where $V$ is 
the fundamental representation of $U(N)$. 
This equation comes from the fact that 
the $U(N)$ action
commutes with the action of the Brauer algebra on the tensor product. 
In the equation, 
$\gamma$ is an irreducible representation of 
the Brauer algebra and $U(N)$. 
It is given by a pair of two Young diagrams 
which have $m-k$ boxes and $n-k$ boxes, 
and $k$ is an integer satisfying $0\le k\le min(m,n)$. 
Taking this into account, the sum of $\gamma$ 
can be 
re-grouped to be 
\begin{eqnarray}
V^{\otimes m}
\otimes \bar{V}^{\otimes n}
&=&
\bigoplus_{k}\left(
\bigoplus_{\gamma_+ \vdash (m-k),\gamma_- \vdash (n-k)}
V_{(\gamma_+,\gamma_-)}^{U(N)}
\otimes V_{(\gamma_+,\gamma_-)}^{B_N(m,n)}\right).
\end{eqnarray} 
The Brauer algebra 
contains the group algebra of the symmetric group 
$S_m \times S_n$ as a subalgebra:
\begin{eqnarray}
V_{\gamma}^{B_N(m,n)}
=\bigoplus_{\alpha \vdash m,\beta \vdash n}
V_{\gamma \rightarrow (\alpha,\beta)} 
\otimes V_{\alpha}^{\mathbb{C}[S_m]}\otimes V_{\beta}^{\mathbb{C}[S_n]}.
\label{Brauer_symmetric}
\end{eqnarray} 
The vector space 
$V_{\gamma \rightarrow (\alpha,\beta)}$ 
accounts for multiplicities in the restriction. 
The dimension of the space, 
$
M^{\gamma}_{(\alpha,\beta)}:=
DimV_{\gamma \rightarrow (\alpha,\beta)}$, 
counts the number of times the irreducible representation 
$(\alpha,\beta)$ appears 
in the irreducible representation $\gamma$ by the formula 
\begin{eqnarray}
M^{\gamma}_{(\alpha,\beta)}
=\sum_{\delta\vdash k}
g(\delta,\gamma_+;\alpha)
g(\delta,\gamma_-;\beta).
\end{eqnarray}
When $k=0$, we have 
\begin{eqnarray}
V_{(\alpha,\beta)}^{B_N(m,n)}
=
V_{\alpha}^{\mathbb{C}[S_m]}\otimes V_{\beta}^{\mathbb{C}[S_n]}. 
\end{eqnarray}


\section{The conserved charges}
\setcounter{equation}{0}
\label{sec:algebra_conservedcharge}

In this appendix, we present some sets of commuting conserved 
charges for the sector 
with excitations by $B_{a}^{\dagger}$ $(a=1,2,3)$. 
The extension to more oscillators is straightforward. 
We will use 
(\ref{building_blocks}) as building blocks.
It is trivial to find that 
the following operators commute each other 
\begin{eqnarray}
tr((G_{L1}^B)^p), \quad tr((G_{L2}^B)^p),\quad tr((G_{L3}^B)^p),
\quad 
tr((G_{R1}^B)^p), \quad tr((G_{R2}^B)^p),\quad tr((G_{R3}^B)^p).
\end{eqnarray}
We call this set $S$. 
We find that it is helpful to use 
the following formula to show several things
\begin{eqnarray}
[tr(G^p),(G^q)_{ij}]=0, 
\label{formula_charges_commutator}
\end{eqnarray}
where $G$ is an operator satisfying the 
$u(N)$ commutation relation. 
We choose $G=G_{L1}^{B}$, $G_{L2}^{B}+G_{L3}^{B}$, and so on. 
From the formula, we find that 
any charges built from the building blocks commute with 
the charges in $S$.

A set of commuting higher charges is given by 
\begin{eqnarray}
S,
&&
tr((G_{L1}+G_{L2}+G_{L3})^p),\quad tr((G_{L1}+G_{L2})^q),
\nonumber  \\
&&
tr((G_{R1}+G_{R2}+G_{R3})^p), \quad tr((G_{R1}+G_{R2})^q).
\label{restricted_charges}
\end{eqnarray}
This is related to the restricted Schur basis in the sense that 
these charges 
have good actions on 
the restricted Schur basis
as we explicitly show in the next appendix.
We can verify that all of these charges commute each other 
with the help of (\ref{formula_charges_commutator}). 
In stead of $tr((G_{L1}+G_{L2})^q)$, we could put $tr((G_{L1}+G_{L3})^q)$ 
but could not put them together. 

Taking into account the fact that 
$G_{L1}+G_{L2}$ is a $u(N)$ generator, 
we can find that 
\begin{eqnarray}
[(G_{L1}+G_{L2})_{ij},tr(H)]=0,
\label{a_formula_G1G2}
\end{eqnarray}
where $H$ is a polynomial built from $G_{L1}$ and $G_{L2}$. 
With the help of this equation, we find  
the operator $tr(H)$ commutes with all operators in 
(\ref{restricted_charges}). 
But in general the trace of a polynomial of $G_{L1}$ and $G_{L2}$ 
does not commutes with the trace of another polynomial of $G_{L1}$ and $G_{L2}$.  
Similarly, the trace of a polynomial built from $G_{L1}+G_{L2}$ and $G_{L3}$ 
commutes with all operators in 
(\ref{restricted_charges}) 
and $tr(H)$ thanks to 
(\ref{formula_charges_commutator}), 
(\ref{a_formula_G1G2}), and the following
\begin{eqnarray}
[(G_{L1}+G_{L2}+G_{L3})_{ij},tr(I)]=0,
\label{a_formula_G1G2G3}
\end{eqnarray}
where $I$ is a polynomial built from $G_{L1}+G_{L2}$ and $G_{L3}$. 
For example, we can choose 
$H=(G_{L1})^{2}(G_{L2})$, 
$I=(G_{L1}+G_{L2})^2(G_{L3})$, and we also have 
$H^{\prime}=(G_{R1})^{2}(G_{R2})$, 
$I^{\prime}=(G_{R1}+G_{R2})^2(G_{R3})$. 
These charges would be relevant for the multiplicity indices \cite{0807.3696}.

\quad 

Another set of commuting higher charges is 
\begin{eqnarray}
S,
&&
tr((G_{L1}+G_{L2}+G_{R3})^p),\quad tr((G_{L1}+G_{L2})^q),
\nonumber  \\
&&
tr((G_{R1}+G_{R2}+G_{L3})^p), \quad tr((G_{R1}+G_{R2})^q).
\label{Brauer_charges}
\end{eqnarray}
This is closely related to the Brauer basis $B_{N}(n_1+n_2,n_3)$.  
We may put $tr((G_{L1}+G_{R3})^q)$ and $tr((G_{R1}+G_{L3})^q)$ 
instead of $tr((G_{L1}+G_{L2})^q)$ and $tr((G_{R1}+G_{R2})^q)$ 
in (\ref{Brauer_charges}), but we cannot put all of these at the same time. 
We are allowed to 
include more charges, for example, 
$tr((G_{L1})^{2}(G_{L2}))$, 
$tr((G_{L1}+G_{L2})^2(G_{R3}))$,  
$tr((G_{R1})^{2}(G_{R2}))$, and 
$tr((G_{R1}+G_{R2})^2(G_{L3}))$.

\quad 

By a similar construction to the previous two cases, 
we can form 
a set of commuting charges that includes
\begin{eqnarray}
&&
S,\quad tr((G_{L1}+G_{L2}+G_{R1})^p)
,\quad tr((G_{R2}+G_{R3}+G_{L3})^p).
\label{new_charges}
\end{eqnarray}
These charges are simultaneously diagonalised 
by 
operators of the form 
$(P^{\gamma}_{\vec{r},ij})_{IKJ}^{MLN}X^I_J Y^{K}_L Z^T{}^M_N$ 
in terms of the Brauer algebra. 
But 
considering carefully, 
we realise that 
such operators do not form  
a complete set of local gauge invariant operators
\footnote{
The following is an exception.
Consider $m=n=1$ in the two-matrix system. 
We have two gauge invariant operators, 
$trXY$ and $trXtrY$. 
From these operators, we 
can find some combinations which have diagonal free two-point functions: 
(1) ($trXtrY+trXY$, $trXtrY-trXY$), 
(2) ($trXtrY-(1/N) trXY$, $trXY$), 
(3) ($trXY-\frac{1}{N}trXtrY$, $trXtrY$).
The first one is the restricted Schur basis, while 
the second one is the Brauer basis. 
The last one is the new basis 
related to the charges (\ref{new_charges}). 
}, in general, 
though free two-point functions are diagonalised. 


\section{Action of the conserved charges}
\setcounter{equation}{0}
\label{Casimirs_action}

In this section, we display eigenvalues of the charges 
given in appendix \ref{sec:algebra_conservedcharge}. 
For more detail see \cite{0807.3696}.

The Schur polynomial basis relevant to the 1/2 BPS sector is given by 
\begin{eqnarray}
|R\rangle = 
tr_{n}(p_{R}B_1^{\dagger \otimes n})|0\rangle
=
\frac{d_{R}}{n!}\sum_{\sigma \in S_n}\chi_{R}(\sigma^{-1} )
tr_{n}(\sigma B_1^{\dagger \otimes n})|0\rangle, 
\end{eqnarray}
where 
$R$ is a Young diagram with $n$ boxes ($c_1(R)\le N$).  
This basis has the following actions
\begin{eqnarray}
&&
tr((B_1^{\dagger}B_1)^p)|R\rangle 
=
C_p(R)|R\rangle ,\nonumber \\
&&
tr((B_1 B_1^{\dagger})^p)|R\rangle 
=
C_p(R)|R\rangle . 
\label{higher_charges_schur}
\end{eqnarray}
$C_p(R)$ is the $p$-th Casimir.

For the restricted Schur basis
\begin{eqnarray}
|R,\vec{r},ij\rangle =tr_{n_1+n_2+n_3}(P^{R}_{\vec{r},ij}
(B_1^{\dagger})^{\otimes n_1}\otimes (B_2^{\dagger})^{\otimes n_2}
\otimes (B_3^{\dagger})^{\otimes n_3}
) 
|0\rangle, 
\end{eqnarray}
we have the following actions of 
the commuting conserved charges 
\begin{eqnarray}
&&tr((G_{L1})^p)|R,\vec{r},ij\rangle=C_p(r_1)|R,\vec{r},ij\rangle, 
\nonumber 
\\
&&tr((G_{L2})^p)|R,\vec{r},ij\rangle=C_p(r_2)|R,\vec{r},ij\rangle, 
\nonumber 
\\
&&tr((G_{L3})^p)|R,\vec{r},ij\rangle=C_p(r_3)|R,\vec{r},ij\rangle, 
\nonumber 
\\
&&tr((G_{L1}+G_{L2}+G_{L3})^p)|R,\vec{r},ij\rangle=C_p(R)|R,\vec{r},ij\rangle,
\nonumber 
\\
&&tr((G_{L1}+G_{L2})^p)|R,\vec{r},ij\rangle=
\sum_{\alpha \vdash (n_1+n_2)}
C_p(\alpha)
\frac{1}{d_{\vec{r}}}
\chi_{\vec{r},jl}^R (P^{\alpha}\circ 1)
|R,\vec{r},il\rangle,
\nonumber 
\\
&&tr((G_{R1}+G_{R2})^p)|R,\vec{r},ij\rangle=
\sum_{\alpha \vdash (n_1+n_2)}C_p(\alpha)
\frac{1}{d_{\vec{r}}}
\chi_{\vec{r},ki}^R (P^{\alpha}\circ 1)
|R,\vec{r},kj\rangle.
\end{eqnarray}
The last two are derived by inserting 
$1=\sum_{\alpha}P^{\alpha}$ and using an equation similar to
(\ref{completeness_equation}).

As was mentioned around 
(\ref{a_formula_G1G2})
(\ref{a_formula_G1G2G3}), 
we have more charges. 
Let us choose
$tr((G_{L1}+G_{L2})^2 G_{L3})$ and 
$tr(G_{L1}^{2} G_{L2})$ for instance. 
The exact eigenvalues have not been clear, 
but it was guessed in \cite{0807.3696} that those will be measuring  
the multiplicity indices. 
Our guess is that the first one 
is related to the restriction 
$S_{n_1+n_2+n_3}\rightarrow S_{n_1+n_2}\times S_{n_3}$
and the second one is related to the restriction
$S_{n_1+n_2}\rightarrow S_{n_1}\times S_{n_2}$. 

\quad 

Likewise for the Brauer basis 
\begin{eqnarray}
|\gamma,\vec{r},ij\rangle =tr_{n_1+n_2,n_3}(Q^{\gamma}_{\vec{r},ij}
(B_1^{\dagger})^{\otimes n_1}\otimes (B_2^{\dagger})^{\otimes n_2}
\otimes (B_3^{\dagger})^{
T \otimes n_3}
) |0\rangle, 
\end{eqnarray}
we have 
\begin{eqnarray}
&&tr((G_{L1})^p)|\gamma,\vec{r},ij\rangle=C_p(r_1)|\gamma,\vec{r},ij\rangle, 
\nonumber 
\\
&&tr((G_{L2})^p)|\gamma,\vec{r},ij\rangle=C_p(r_2)|\gamma,\vec{r},ij\rangle, 
\nonumber 
\\
&&tr((G_{L3})^p)|\gamma,\vec{r},ij\rangle=C_p(r_3)|\gamma,\vec{r},ij\rangle, 
\nonumber 
\\
&&tr((G_{L1}+G_{L2}+G_{R3})^p)|\gamma,\vec{r},ij\rangle=C_p(\gamma)|\gamma,\vec{r},ij\rangle,
\nonumber 
\\
&&tr((G_{L1}+G_{L2})^p)|\gamma,\vec{r},ij\rangle=
\sum_{\alpha \vdash (n_1+n_2)}C_p(\alpha)
\frac{1}{d_{\vec{r}}}
\chi_{\vec{r},jl}^{\gamma} (P^{\alpha}\circ 1)
|\gamma,\vec{r},il\rangle,
\nonumber 
\\
&&tr((G_{R1}+G_{R2})^p)|\gamma,\vec{r},ij\rangle=
\sum_{\alpha \vdash (n_1+n_2)}C_p(\alpha)
\frac{1}{d_{\vec{r}}}
\chi_{\vec{r},ki}^{\gamma} (P^{\alpha}\circ 1)
|\gamma,\vec{r},kj\rangle,
\nonumber 
\\
&&tr((G_{L1}+G_{R3})^p)|\gamma,\vec{r},ij\rangle=
\sum_{\gamma_1}C_p(\gamma_1)
\frac{1}{d_{\vec{r}}}
\chi_{\vec{r},jl}^{\gamma} (P^{\gamma_1})
|\gamma,\vec{r},il\rangle,
\nonumber 
\\
&&tr((G_{R1}+G_{L3})^p)|\gamma,\vec{r},ij\rangle=
\sum_{\gamma_1}C_p(\gamma_1)
\frac{1}{d_{\vec{r}}}
\chi_{\vec{r},ki}^{\gamma} (P^{\gamma_1})
|\gamma,\vec{r},kj\rangle.
\nonumber 
\end{eqnarray}
In the last two equations, 
$\gamma_1$ runs over irreducible representations of $B_N(n_1,n_3)$.  
The 5th and 6th equations 
take the following forms 
at $k=0$ 
\begin{eqnarray}
&&tr((G_{L1}+G_{L2})^p)|(\gamma_+,\gamma_-,k=0),\vec{r},ij\rangle=
C_p(\gamma_+)|(\gamma_+,\gamma_-,k=0),\vec{r},ij\rangle,
\nonumber 
\\
&&tr((G_{R1}+G_{R2})^p)|(\gamma_+,\gamma_-,k=0),\vec{r},ij\rangle=
C_p(\gamma_+)|(\gamma_+,\gamma_-,k=0),\vec{r},ij\rangle.
\end{eqnarray}


\section{Two special sectors of the Brauer basis}
\setcounter{equation}{0}
\label{sec:specialsectors_Brauer}

In this appendix, we will show explicit forms of 
the operator in the two sectors
where the integer $k$ 
takes the minimum possible value and the maximum possible value. 

\subsection{$k=0$}

When $k$ takes zero, 
we have $\gamma_-=r_3$ and 
the multiplicity indices on the $k=0$ operators 
run over from $1$ to 
\begin{eqnarray}
M_{\vec{r}}^{\gamma(k=0)}
=
g(r_1,r_2;\gamma_+).
\end{eqnarray} 
In this sector, some 
special properties are available to rewrite the form of 
the operator $Q^{\gamma(k=0)}_{\vec{r},ij}$
\cite{0709.2158}: 
\begin{eqnarray}
Q^{\gamma(k=0)}_{\vec{r},ij}
&=&
Dim\gamma \sum_b 
\chi^{\gamma}_{\vec{r},ij}(b)b^{\ast}
\nonumber \\
&=&
Dim\gamma 
\sum_{b\in \mathbb{C}[S_{n_1+n_2} \times S_{n_3}]} 
\chi^{\gamma}_{\vec{r},ij}(b)b^{\ast}
\nonumber \\
&=&
Dim\gamma \sum_{\sigma \in S_{n_1+n_2}}
\sum_{\tau \in S_{n_3}}  
\chi^{\gamma_+}_{(r_1,r_2),ij}(\sigma)
\chi_{r_3}(\tau)
1^{\ast} (\sigma \circ \tau)^{-1}
\nonumber \\
&=&
Dim\gamma 
\frac{(n_1+n_2) !}{d_{\gamma_{+}}}
\frac{n_3 !}{d_{r_3}}
1^{\ast}
P^{\gamma_{+}}_{(r_1,r_2),ij} p_{r_3},
\end{eqnarray}
where 
\begin{eqnarray}
P^{\gamma_{+}}_{(r_1,r_2),ij}  =
\frac{d_{r_{+}}}{(n_1+n_2) !}
\sum_{\sigma \in S_{n_1+n_2}}
\chi^{\gamma_{+}}_{(r_1,r_2),ij}(\sigma)\sigma^{-1}. 
\end{eqnarray}
$1^{\ast}$ is a specific element 
expressed by a linear combination of 
elements 
in the Brauer algebra \cite{0709.2158}. 
Using a formula of $1^{\ast}$, we have
\begin{eqnarray}
&&
tr_{n_1+n_2,n_3}(Q^{\gamma(k=0)}_{\vec{r},ij} 
X\otimes Y \otimes Z^{T}
)
\nonumber \\
&=&
Dim\gamma 
\frac{(n_1+n_2) !}{d_{\gamma_{-}}}
\frac{n_3 !}{d_{r_3}}
\frac{1}{N^{n}}
tr_{n}(\Omega_{n}^{-1}
P^{\gamma_{+}}_{(r_1,r_2),ij} p_{r_3}
X\otimes Y\otimes Z
), 
\end{eqnarray}
where $n\equiv n_1+n_2+n_3$.
Note that $Z$'s are not transposed. 
It is an exercise to  
express it as a linear combination of 
restricted Schur polynomials. Making use of  
\begin{eqnarray}
\Omega_{n}^{-1}
P^{\gamma_{+}}_{(r_1,r_2),ij} p_{r_3}
&=&
\sum_{S,\vec{s},kl}
\frac{1}{d_s}
\chi_{s,kl}^S
\left(
\Omega_{n}^{-1}
P^{\gamma_{+}}_{(r_1,r_2),ij}p_{r_3}
\right)
P^S_{s,kl} 
\nonumber \\
&=&
\sum_{S,kl}
\frac{1}{d_{\vec{r}}}
\chi_{\vec{r},kl}^S
\left(P^{\gamma_{+}}_{(r_1,r_2),ij}
\right)
\frac{N^n}{n!}
\frac{d_{S}}{DimS}
P^S_{\vec{r},kl} ,
\end{eqnarray}
where the second line is obtained by using 
\begin{eqnarray}
\Omega_{n}^{-1}
=\frac{N^n}{n!}
\sum_{S\vdash n}\frac{d_S}{DimS}P^S ,
\end{eqnarray}
we obtain 
\begin{eqnarray}
&&
tr_{n_1+n_2,n_3}(Q^{\gamma(k=0)}_{\vec{r},ij} 
X\otimes Y \otimes Z^{T}
)
\nonumber \\
&=&
Dim\gamma 
\frac{(n_1+n_2) !}{d_{r_{+}}}
\frac{n_3 !}{d_{r_3}}
\frac{1}{N^{n}}
tr_{n}(\Omega_{n}^{-1}
P^{\gamma_{+}}_{(r_1,r_2),ij}p_{r_3} 
X\otimes Y\otimes Z
)
\nonumber \\
&=&
Dim\gamma 
\frac{(n_1+n_2) !n_3 !}{n!}
\frac{1}{d_{r_{+}}d_{r_3}}
\sum_{S,kl}
\frac{1}{d_{\vec{r}}}
\chi_{\vec{r},kl}^S
\left(P^{\gamma_{+}}_{(r_1,r_2),ij}
\right)
\frac{d_S}{DimS}
tr_{n}(
P^S_{\vec{r},kl} X\otimes Y\otimes Z
).
\end{eqnarray}


\subsection{$k=n_1+n_2=n_3$}

We consider the case $n_1+n_2=n_3$ for simplicity. 
When the integer $k$ takes the maximum possible value
$k=n_1+n_2=n_3$, $\gamma=(\emptyset,\emptyset)$. 
The multiplicity is given by 
\begin{eqnarray}
M_{\vec{r}}^{\gamma}
=
g(r_1,r_2;r_3).
\end{eqnarray}

\quad

First introduce the contraction operator \cite{0911.4408}
\begin{eqnarray}
C_{(k)}=\sum_{\sigma \in S_{k}}
\sigma C_{1\bar{1}}
\cdots C_{k\bar{k}}\sigma^{-1},
\end{eqnarray}
which satisfies 
\begin{eqnarray}
C_{(k)}^2
=
N^k \Omega_k
C_{(k)} .
\end{eqnarray}
In terms of this, we can construct the operator
\begin{eqnarray}
Q^{\gamma (k=n_1+n_2=n_3)}_{\vec{r},ij}
=\frac{d_{r_3}}{Dim r_3 n_3!}
(P^{r_3}_{(r_1,r_2),ij} \otimes 1)C_{(k)}.
\end{eqnarray}
We will check it satisfies the following required relations 
\begin{eqnarray}
&& Q_{\vec{r},ij}
Q_{\vec{s},kl}
=\delta_{\vec{r}\vec{s}}\delta_{jk}Q_{\vec{r},il},
\nonumber \\
&&
tr_{n_1+n_2,n_3}(Q^{\gamma (k=n_1+n_2=n_3)}_{\vec{r},ij})
=d_{\vec{r}} \delta_{ij}.
\end{eqnarray}
The proof of the first equation is as follows: 
\begin{eqnarray}
&&
(P^{r_3}_{(r_1,r_2),ij}\otimes 1)C_{(k)}
(P^{s_3}_{(s_1,s_2),kl}\otimes 1)C_{(k)}
\nonumber \\
&=&
\delta_{\vec{r}\vec{s}}\delta_{jk}
(P^{r_3}_{(r_1,r_2),il}\otimes 1)C_{(k)}^2
\nonumber \\
&=&
\delta_{\vec{r}\vec{s}}\delta_{jk}
N^k\Omega_k
(P^{r_3}_{(r_1,r_2),ij}\otimes 1)C_{(k)}
\nonumber \\
&=&
\delta_{\vec{r}\vec{s}}\delta_{jk}
\frac{Dim r_3 n_3!}{d_{r_3}}
(P^{r_3}_{(r_1,r_2),ij}\otimes 1)C_{(k)}. 
\end{eqnarray}
We can also verify the second equation as 
\begin{eqnarray}
&&
tr_{n_1+n_2,n_3}(
Q^{\gamma (k=n_1+n_2=n_3)}_{\vec{r},ij}
)
\nonumber \\
&=&
\frac{d_{r_3}}{Dim r_3 n_3!}
\sum_{\sigma\in S_k}
tr_{n_1+n_2}(
\sigma 
P^{r_3}_{(r_1,r_2),ij}\sigma^{-1}
)
\nonumber \\
&=&
\frac{d_{r_3}}{Dim r_3 } 
tr_{n_1+n_2}(
P^{r_3}_{(r_1,r_2),ij}
)
\nonumber \\
&=&
\frac{d_{r_3}}{Dim r_3 } 
Dim r_3 d_{r_1}d_{r_{2}} \delta_{ij}
\nonumber \\
&=&
d_{r_{1}}
d_{r_2}d_{r_{3}}\delta_{ij}. 
\end{eqnarray}

The operator looks like 
\begin{eqnarray}
&&
tr_{n_1+n_2,n_3}(
Q^{\gamma (k=n_1+n_2=n_3)}_{(r_1,r_2,r_3),ij}
X^{\otimes n_1}\otimes Y^{\otimes n_2} \otimes Z^{T\otimes n_3}
)
\nonumber \\
&=&
\frac{d_{r_3}}{Dim r_3 n_3!}
tr_{n_1+n_2,n_3}(
(P^{r_3}_{(r_1,r_2),ij}\otimes 1)C_{(k)}
X^{\otimes n_1}\otimes Y^{\otimes n_2} \otimes Z^{T\otimes n_3}
)
\nonumber \\
&=&
\frac{d_{r_3}}{Dim r_3 }
tr_{n_1+n_2}(
 P^{r_3}_{(r_1,r_2),ij}
(ZX)^{\otimes n_1} \otimes (ZY)^{\otimes n_2} 
). 
\end{eqnarray}
This is nothing but 
a restricted Schur polynomial built from  
the two matrices $ZX$ and $ZY$.


\section{The branching rule of the Brauer algebra}
\setcounter{equation}{0}
\label{ref:derivationofkk1k2}

The branching rule 
for $B_N(m_1,n_1)\times B_N(m_2,n_2) \subseteq B_N(m,n)$ is 
given by 
\begin{eqnarray}
&&
V_{\gamma}
=\bigoplus_{\gamma_1,\gamma_2}M_{\gamma_1,\gamma_2}^{\gamma}
V_{\gamma_1}\otimes V_{\gamma_2},
\end{eqnarray}
where $\gamma_1$, $\gamma_2$, and $\gamma$ 
are irreducible representations of 
$B_N(m_1,n_1)$, $B_N(m_2,n_2)$ and $B_N(m,n)$, respectively. 
The multiplicity
$M_{\gamma_1,\gamma_2}^{\gamma}$ 
is expressed in terms of 
the Littlewood-Richardson 
coefficient as \cite{Koike1989,Halverson1996}
\begin{eqnarray}
&&\hspace{-0.8cm}
M_{\gamma_1,\gamma_2}^{\gamma}
=
\nonumber \\
&& \hspace{-0.8cm}
\sum_{\rho,\zeta,\theta,\kappa}
\left(
\sum_{\delta}
g(\delta,\rho;\gamma_{1+})
g(\delta,\zeta;\gamma_{2-})
\right)
\left(
\sum_{\epsilon}
g(\epsilon,\theta;\gamma_{1-})
g(\epsilon,\kappa;\gamma_{2+})
\right)
g(\rho,\kappa;\gamma_{+})
g(\zeta,\theta;\gamma_{-}).
\end{eqnarray}
Denote the number of boxes contained in 
a Young diagram $\alpha$ by $n(\alpha)$. 
From consistency conditions for the Littlewood-Richardson coefficients,
we need the following conditions
\begin{eqnarray}
&&
n(\delta)+n(\rho)=m_1-k_1,
\nonumber \\
&&
n(\delta)+n(\zeta)=n_2-k_2,
\nonumber \\
&&
n(\epsilon)+n(\theta)=n_1-k_1,
\nonumber \\
&&
n(\epsilon)+n(\kappa)=m_2-k_2,
\nonumber \\
&&
n(\rho)+n(\kappa)=m-k,
\nonumber \\
&&
n(\zeta)+n(\theta)=n-k.
\label{severalconditions_Box}
\end{eqnarray}
We also have 
\begin{eqnarray}
m_1+m_2=m, \quad n_1+n_2=n. 
\end{eqnarray}
From these conditions, we have 
\begin{eqnarray}
n(\delta)+n(\epsilon)=k-(k_1+k_2). 
\label{delta_epsilon_k}
\end{eqnarray}
For this to be satisfied for any Young diagrams 
$\delta$, $\epsilon$, one should have 
\begin{eqnarray}
k-(k_1+k_2) \ge 0. 
\end{eqnarray}
The equality is if and only if $n(\delta)=0$ and $n(\epsilon)=0$, 
which is the case 
$m_1-k_1+m_2-k_2=m-k$ and 
$n_1-k_1+n_2-k_2=n-k$. 
The multiplicity becomes for this case
\begin{eqnarray}
M_{\gamma_1,\gamma_2}^{\gamma (k=k_1+k_2)}
=
g(\gamma_{1+},\gamma_{2+};\gamma_{+})
g(\gamma_{1-},\gamma_{2-};\gamma_{-}).
\end{eqnarray}
The $\gamma_+$-sector and the $\gamma_-$-sector are completely 
decoupled. We call this form factorised form.




\end{document}